\documentclass[sigconf,balance=false]{acmart}

\usepackage{popets}
\setcopyright{popets}
\copyrightyear{2025}
\acmYear{2025}
\acmVolume{2025}
\acmNumber{4}
\acmDOI{}
\acmISBN{}
\acmConference{(Preprint) Proceedings on Privacy Enhancing Technologies}
\usepackage{algorithm}
\usepackage{algorithmic}
\usepackage{cleveref}
\usepackage{tikz} 
\usepackage[shortcuts,acronym]{glossaries}
\usepackage[most]{tcolorbox}
\usepackage{xspace}
\usepackage{multirow}
\usepackage{dsfont}
\usepackage{url}
\usepackage{subfig}
\usetikzlibrary{matrix} 

\newacronym{GDPR}{GDPR}{General Data Protection Regulation}
\newacronym{CNN}{CNN}{Convolutional Neural Network}
\newacronym{HIPAA}{HIPAA}{Health Insurance Portability and Accountability Act}
\newacronym{ML}{ML}{Machine Learning}
\newacronym{FL}{FL}{Federated Learning}
\newacronym[plural=DTs,firstplural=Decision Trees (DTs)]{DT}{DT}{Decision Tree}
\newacronym[plural=GDBTs]{GBDT}{GBDT}{Gradient Boosting Decision Trees}
\newacronym{XGBoost}{XGBoost}{eXtreme Gradient Boosting}
\newacronym{PS}{PS}{Parameter Server}
\newacronym{HE}{HE}{Homomorphic Encryption}
\newacronym{DP}{DP}{Differential Privacy}
\newacronym{LDP}{LDP}{Local Differential Privacy}
\newacronym{MPC}{MPC}{Secure Multi-Party Computation}
\newacronym{LSH}{LSH}{Locality-Sensitive Hashing}
\newacronym[plural=ANNs,firstplural=Artificial Neural Networks (ANNs)]{ANN}{ANN}{Artificial Neural Network}
\newacronym{GAN}{GAN}{Generative Adversarial Network}
\newacronym{SOTA}{SotA}{State-of-the-Art}
\newacronym{AUC}{AUC-ROC}{Area Under the ROC Curve}
\newacronym{ROC}{ROC}{Receiver Operating Characteristic}
\newacronym{HFL}{HFL}{Horizontal Federated Learning}
\newacronym{IoT}{IoT}{Internet of Things}
\newacronym[plural=RFs,firstplural=Random Forests (RFs)]{RF}{RF}{Random Forest}
\newacronym{iid}{IID}{identically and independently distributed}
\newacronym{RA}{RA}{Reconstruction Accuracy}
\newacronym{MILP}{MILP}{Mixed-Integer Linear Programming}
\newacronym{OvR}{OvR}{One-vs-Rest}
\glsdisablehyper

\makeglossaries

\newcommand{\mypar}[1]{\addvspace{1.8pt}\noindent\textbf{#1.}\xspace}

\newcommand{\name}{TimberStrike\xspace}

\newtcolorbox[auto counter, number within=section]{answer}{%
  colback=gray!10, 
  colframe=black, 
  toprule=-0.25mm, 
  bottomrule=-0.25mm, 
  leftrule=1.5mm, 
  rightrule=-0.25mm, 
  left=0.1mm, right=0.1mm, top=0.1mm, bottom=0.1mm, 
  enhanced,
  arc=0mm,
  breakable = true,
}

\newtcolorbox[auto counter, number within=section]{definition}{%
  colback=gray!10, 
  colframe=orange, 
  toprule=-0.25mm, 
  bottomrule=-0.25mm, 
  leftrule=1.5mm, 
  rightrule=-0.25mm, 
  left=0.1mm, right=0.1mm, top=0.1mm, bottom=0.1mm, 
  enhanced,
  arc=0mm,
  breakable = true,
}

\begin{document}

\title[\name]{\name: Dataset Reconstruction Attack Revealing Privacy Leakage in Federated Tree-Based Systems}


\author{Marco Di Gennaro}
\authornote{These authors contributed equally to this research.}
\orcid{0009-0008-1415-4787}
\affiliation{%
  \institution{Politecnico di Milano}
  \city{Milan}
  \state{}
  \country{Italy}}
\email{marco.digennaro@polimi.it}

\author{Giovanni De Lucia}
\authornotemark[1]
\affiliation{%
  \institution{Politecnico di Milano}
  \city{Milan}
  \state{}
  \country{Italy}}
\email{giovanni.delucia@mail.polimi.it}

\author{Stefano Longari}
\orcid{0000-0002-7533-4510}
\affiliation{%
  \institution{Politecnico di Milano}
  \city{Milan}
  \state{}
  \country{Italy}}
\email{stefano.longari@polimi.it}

\author{Stefano Zanero}
\orcid{0000-0003-4710-5283}
\affiliation{%
  \institution{Politecnico di Milano}
  \city{Milan}
  \state{}
  \country{Italy}}
\email{stefano.zanero@polimi.it}

\author{Michele Carminati}
\orcid{0000-0001-8284-6074}
\affiliation{%
  \institution{Politecnico di Milano}
  \city{Milan}
  \state{}
  \country{Italy}}
\email{michele.carminati@polimi.it}

\renewcommand{\shortauthors}{M. Di Gennaro et al.}


\begin{abstract}
  \acrlong{FL} has emerged as a privacy-oriented alternative to centralized \acrlong{ML}, enabling collaborative model training without direct data sharing. While extensively studied for neural networks, the security and privacy implications of tree-based models remain underexplored. This work introduces \name, an optimization-based dataset reconstruction attack targeting horizontally federated tree-based models. Our attack, carried out by a single client, exploits the discrete nature of decision trees by using split values and decision paths to infer sensitive training data from other clients. We evaluate \name on \acrlong{SOTA} federated gradient boosting implementations across multiple frameworks, including Flower, NVFlare, and FedTree, demonstrating their vulnerability to privacy breaches. On a publicly available stroke prediction dataset, \name consistently reconstructs between 73.05\% and 95.63\% of the target dataset across all implementations. We further analyze \acrlong{DP}, showing that while it partially mitigates the attack, it also significantly degrades model performance. Our findings highlight the need for privacy-preserving mechanisms specifically designed for tree-based \acrlong{FL} systems, and we provide preliminary insights into their design.

\end{abstract}


\maketitle

\section{Introduction}

\acrfull{FL}~\cite{mcmahan_communication-efficient_2023} is a \acrfull{ML} paradigm in which decentralized nodes can collaboratively train a model. Specifically, these nodes train a shared \textit{global model} under the coordination of a central server, known as \textit{\acrfull{PS}}, while keeping their local data private. There are different types of \acrlong{FL}. Between them, we focus on \textit{horizontal} \acrshort{FL}~\cite{kumar_impact_2023}, where clients (or nodes) have different datasets but share the same feature space.
\acrshort{FL} is considered an alternative to traditional centralized \acrshort{ML} training in privacy-sensitive domains, such as healthcare~\cite{habehh_machine_2021,murdoch_inevitable_2013}. The adoption of \acrshort{FL} in applications like healthcare~\cite{santos_federated_2024} and finance~\cite{abdul_salam_federated_2024} is largely driven by data-sharing regulations, such as the European Union's \acrfull{GDPR}~\cite{gdpr} and the United States' \acrfull{HIPAA}~\cite{centers_for_medicare__medicaid_services_health_1996}. However, even if the paradigm does not involve data exposure, \acrshort{FL} does not always guarantee the privacy of training data. Indeed, several studies have demonstrated that attackers can infer sensitive information from the exchanged model updates~\cite{hitaj_deep_2017,zhu_deep_2019}, such as individual training samples or specific dataset properties.

Originally designed for \glspl{ANN}, \acrshort{FL} has been extended to tree-based models~\cite{li_fedtree_2023,ma_gradient-less_2023,maddock_federated_2022,tian_federboost_2022}, given their strong performance on tabular data~\cite{lindskog-munzing_treexnets_2024}. In particular, works proposing federated tree-based systems adapt \acrshort{FL} settings to ensembles of \glspl{DT}, such as \acrfull{GBDT}~\cite{friedman_greedy_2001} and \acrfull{XGBoost}~\cite{chen_xgboost_2016}. 

To the best of our knowledge, unlike \glspl{ANN}, for which several privacy attacks and defenses have been proposed for both tabular~\cite{vero_tableak_2023} and non-tabular data~\cite{zhu_deep_2019}, the privacy of tree-based models in \acrshort{FL} remains underexplored. Prior works~\cite{roth_nvidia_2022,li_fedtree_2023} have addressed privacy concerns in federated tree-based systems by adapting existing privacy defenses such as \acrfull{DP}~\cite{dwork_differential_2006, ji_differential_2014}, \acrfull{MPC}~\cite{lindell_secure_2005}, and \acrfull{HE}~\cite{acar_survey_2018}. Additionally, some studies have explored privacy attacks against tree-based \textit{vertical} \acrshort{FL} systems~\cite{chen_fia-te_2024,takahashi_eliminating_2023,luo_feature_2021}. 
In contrast, the \acrfull{SOTA} lacks an in-depth examination of privacy attacks in tree-based horizontal \acrshort{FL} scenarios. This gap in the \acrshort{SOTA} is relevant given the strong appeal of tree-based models for their interpretability and high performance on tabular datasets, which are especially prevalent in healthcare and other critical sectors where the horizontal \acrshort{FL} paradigm is widely used.

In this work, we address this gap by proposing \name, a novel optimization-based dataset reconstruction attack that exploits privacy vulnerabilities in horizontally federated tree-based models. Our primary objective is to demonstrate the vulnerability to reconstruction attacks of the most promising variants of tree-based \acrshort{FL} systems, implemented by well-known frameworks such as Flower~\cite{beutel_flower_2022} and NVFlare~\cite{roth_nvidia_2022}. Regardless of the approach, we aim to demonstrate that the attributes that make tree-based models attractive for \acrshort{FL} systems, such as their discrete split values and explicit decision paths, effectively expose them to privacy leakage in collaborative scenarios. Our attack is formalized and evaluated on federated gradient boosting models (e.g., \acrshort{GBDT} and \acrshort{XGBoost}), as these are supported by several widely adopted frameworks.
We consider a threat model in which the adversary acts as an \textit{honest-but-curious client}, seeking to steal other clients' training data by reconstructing it, without disrupting the training process to remain undetected. Our intuition is that an adversary with access to the trees built by other clients may use the splitting criteria generated from their training datasets to infer them.
The attack consists of two main phases. The first one, namely \textit{First-Tree Probing}, targets the first tree of the victim client and infers fundamental information, such as the number of samples in the training set and the label distribution. By extracting this information and following the tree splits, the adversary can generate an initial version of the reconstructed dataset. The second phase, namely \textit{Feature Range Inference}, refines the feature ranges in the reconstructed dataset, improving the reconstruction of the victim's data by solving an optimization problem for each subsequent victim's tree.
We adapt our attack methodology to four different federated \acrshort{XGBoost} variants and one \acrshort{GBDT} implementation. We consider three approaches implemented in the Flower framework~\cite{beutel_flower_2022} (Bagging, Cyclic, and FedXGBllr~\cite{ma_gradient-less_2023}), the histogram-based implementation in NVFlare~\cite{roth_nvidia_2022}, and a standalone framework called FedTree~\cite{li_fedtree_2023}. Adapting to the diverse aggregation mechanisms and implementation strategies of these systems constitutes a central challenge of our work. Finally, after analyzing \acrshort{SOTA} privacy defenses, we offer insights into how horizontal tree-based \acrshort{FL} systems should be designed to be resilient against reconstruction attacks.

We experimentally evaluate our approach on two healthcare datasets: Stroke Prediction~\cite{mxfb-sc71-23} and Pima Indians Diabetes~\cite{choubey2017classification}. We demonstrate that the \acrfull{SOTA} in horizontal tree-based \acrshort{FL}, implemented in the most popular framework, is vulnerable to the \name attack. Specifically, we show that \name can reconstruct a significant portion of the target dataset, achieving a \acrfull{RA} consistently above 73.05\% on the Stroke dataset across all features and implementations, and up to 95.63\% when considering only the most important features. We further analyze the impact of each attack phase and its dependency on the algorithms' hyperparameters. Additionally, we show that while classical privacy defenses like \acrshort{DP} reduce attack effectiveness, they fail to fully mitigate it and significantly degrade model performance, making the privacy-utility trade-off difficult to manage. Finally, we provide preliminary insights into the design of future privacy-preserving horizontal tree-based \acrshort{FL} systems by examining the information leveraged by our attack.

\mypar{Open Source and Ethics} We release the attack code for all evaluated systems\footnote{\url{https://github.com/necst/TimberStrike}}. Ethical considerations are discussed in~\Cref{appendix:ethical}.

Our main contributions to the \acrshort{SOTA} are the following:
\begin{itemize}
    \item We propose an optimization-based dataset reconstruction attack targeting tree-based horizontal \acrshort{FL} systems. To the best of our knowledge, this is the first work to investigate reconstruction attacks in this specific setting. Our attack allows an honest-but-curious client-side adversary to infer other clients' training data by leveraging the exchanged model updates. Importantly, the proposed method is compatible with several \acrshort{SOTA} tree-based \acrshort{FL} frameworks.
    \item We demonstrate that existing defense mechanisms—when compatible with our threat model—either fail to mitigate the attack or incur a substantial loss in model utility. This highlights the need for robust privacy-preserving mechanisms in tree-based \acrshort{FL} systems.
    \item We provide some preliminary insights into the design principles of an ideal tree-based \acrshort{FL} system that is robust against such reconstruction attacks.
\end{itemize}
\section{Background}\label{sec:background}

In this section, we introduce the main concepts needed to understand our work. Indeed, we discuss the primers on \acrfull{GBDT}, \acrfull{XGBoost}, \acrlong{FL}, and federated tree-based systems. For the latter, we describe the \acrshort{SOTA} implementations for which we design a dataset reconstruction attack in this work. In addition, we provide background on \acrlong{DP} in~\Cref{appendix:background}.

\subsection{Primer on GBDT and XGBoost} \label{sub:xgboostgdbt}

To understand our dataset reconstruction attack, we first establish the key concepts of \acrshort{XGBoost} and gradient boosting.

\mypar{\acrfullpl{DT}} A \acrshort{DT}~\cite{maimon_decision_2005} is a tree-based model used for \textit{classification} and \textit{regression} tasks. It consists of \textit{nodes} representing decision rules, \textit{branches} representing possible outcomes, and \textit{leaves} containing final predictions. The construction of a \acrshort{DT} involves recursively splitting the training dataset based on feature values to maximize information gain or minimize impurity measures such as the Gini index or entropy. The final output of a \acrshort{DT} is determined by the values in its leaf nodes (\textit{leaf values}).

\subsubsection{\acrlong{GBDT} and \acrshort{XGBoost}} \acrshort{GBDT}~\cite{friedman_greedy_2001} is an ensemble learning method that trains \acrshortpl{DT} sequentially, with each new tree correcting the residual errors of the previous ones. After training all trees, the \textit{learning rate} determines each tree's contribution. \acrshort{XGBoost}~\cite{chen_xgboost_2016} is an optimized implementation of \acrshort{GBDT} that improves training efficiency by incorporating \( L_1 \) and \( L_2 \) regularization, as well as parallelizing tree construction and tree pruning.  Below, we introduce key definitions related to \acrshort{XGBoost} and \acrshort{GBDT}.

\mypar{Base Score and Initial Predictions} The \textit{base score} is a global model parameter. It is the global bias of the model, i.e., the initial prediction before any trees are trained. This value is crucial in \name since it influences Hessian calculations in the first trained tree, which we exploit for dataset reconstruction.

\mypar{Gradient and Hessian Computation} Each data point contributes to training a new tree through the gradient and Hessian values that are then used to compute the gain for a certain split:
\begin{equation}
\small
g_i = \partial_{\hat{y_i}^{(t-1)}} l(y_i, \hat{y_i}^{(t-1)}), \quad h_i = \partial_{\hat{y_i}^{(t-1)}}^2 l(y_i, \hat{y_i}^{(t-1)}),
\label{eq:hessian_gradient}
\end{equation}
where $l$ is the loss function, $y_i$ is the true label, $\hat{y_i}^{(t-1)}$ is the prediction of the $i$-th sample at iteration $t-1$, $g_i$ (gradient from sample $i$) measures how much a sample's prediction needs to be corrected and $h_i$ (Hessian value of sample $i$) determines confidence in the correction. 
Finally, the total Hessian value and gradient at a node are respectively $H = \sum_{i=1}^{N} h_i$ and $G = \sum_{i=1}^{N} g_i$
where $N$ is the number of samples assigned to that node.

\mypar{Histogram-Based Hessian computation} It is possible to use histogram-based optimization for split identification. For each feature, continuous feature values are discretized into histogram bins, reducing memory usage and computational cost. A small number of split points is proposed. The algorithm accumulates the gradient and Hessian values within each bin, enabling rapid computation of split gains and improving the efficiency of tree construction.

\mypar{Tree Construction and Leaf Assignments} \acrshort{XGBoost} trees are built using a histogram-based approach that selects the best split by maximizing the gain:
\begin{equation} 
\small
\label{eq:xgboost_split_gain}
\text{Gain} = \frac{1}{2} \left( \frac{G_L^2}{H_L + \lambda} + \frac{G_R^2}{H_R + \lambda} - \frac{(G_L + G_R)^2}{H_L + H_R + \lambda} \right) - \gamma,
\end{equation}
where $G_L$, $H_L$, $G_R$, and $H_R$ are total gradients and Hessians for the left and right child nodes, while $\lambda$ and $\gamma$ are regularization parameters. At inference time, each sample follows a decision path dictated by its feature values, eventually landing in a leaf node.

\subsection{Federated Learning} \label{sub:fl_background}
\acrfull{FL}~\cite{mcmahan_communication-efficient_2023} is a learning paradigm in which multiple parties collaboratively train an \acrshort{ML} model while keeping data decentralized, ensuring it remains on client devices.
\acrshort{FL} typically adopts a \textit{client-server} architecture, relying on a central server, commonly referred to as the \acrfull{PS}. Clients train a model on their local datasets and send the resulting updates to the server. The server then aggregates these updates and returns an updated global model, which the clients use for subsequent training rounds or predictions. Fully decentralized architectures~\cite{camajori_tedeschini_decentralized_2022}, where clients exchange and aggregate model updates with peers, have also emerged, but are beyond the scope of this work.

\mypar{Federated Learning Classification} 
\acrshort{FL} can be classified into three categories~\cite{kumar_impact_2023,liu_threats_2022}: \textit{horizontal}, \textit{vertical}, and federated \textit{transfer learning}. In \textit{horizontal} \acrshort{FL}, clients share the same feature space but have different row samples. In \textit{vertical} \acrshort{FL}, clients have the same row samples but different feature spaces. Federated \textit{transfer learning}, on the other hand, occurs when clients differ in both feature spaces and row samples. In this work, we target horizontal \acrshort{FL} systems.

\subsection{Federated Tree-Based Systems}\label{sub:fltrees_background}

Recent studies demonstrate that \acrshort{FL} can be applied to tree-based models~\cite{chatel_sok_2021, yamamoto_new_2020, yamamoto_efl-boost_2022, wang_decision_2024}. For instance, researchers proposed \acrshort{FL} systems based on models like \acrfull{GBDT}~\cite{li_fedtree_2023}, \acrshort{XGBoost}~\cite{yang_tradeoff_2019}, and \acrfull{RF}~\cite{park_random_2023, hauschild_federated_2022}. In this work, we focus on several federated gradient boosting implementations integrated into widely adopted frameworks, including Flower Bagging and Cyclic~\cite{beutel_flower_2022}, NVFlare's histogram aggregation~\cite{roth_nvidia_2022}, FedXGBllr~\cite{ma_gradient-less_2023}, and FedTree~\cite{li_fedtree_2023}. Our focus is driven by the unique vulnerabilities of \acrshort{GBDT} in the federated setting. Indeed, \acrshort{GBDT} learns trees sequentially by leveraging detailed gradient statistics to refine prior errors, creating a richer and more exploitable attack surface for data reconstruction. Since our attack targets these specific mechanisms, \acrshortpl{RF} and simpler \acrshortpl{DT} are outside our work scope.

\begin{figure}[t]
    \centering
    \includegraphics[width=0.8\linewidth]{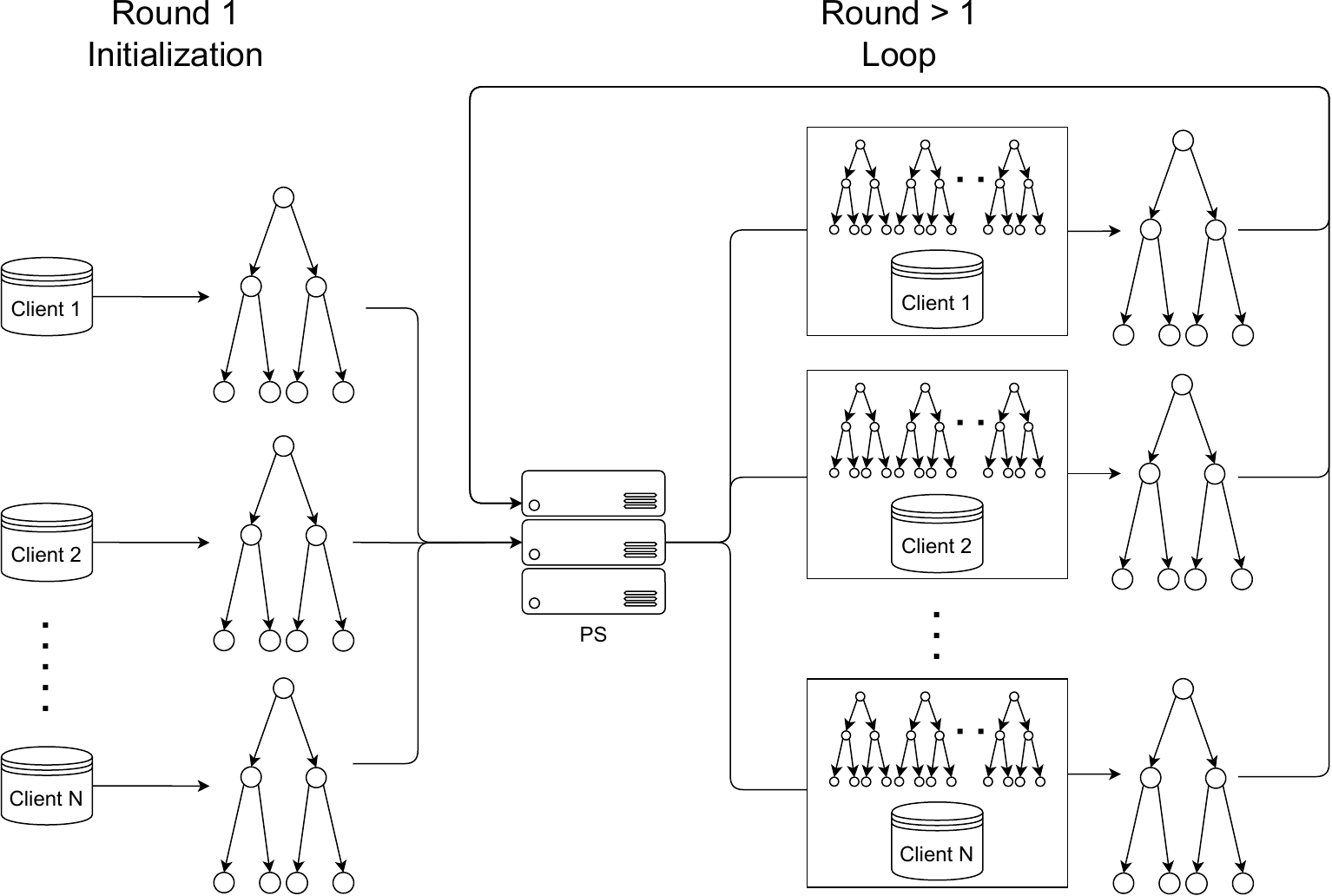}
    \caption{Flower Bagging algorithm schema.}
    \label{fig:flower_bagging}
    \Description{Diagram showing the Flower bagging algorithm flow: multiple clients train local XGBoost models, which are then sent to the server for aggregation through concatenation.}
\end{figure}

\mypar{Flower XGBoost Bagging}
\label{sub:xgboost_bagging} The server distributes the global model to each client in the environment. The clients then use this global model as a starting point for the next boosting round. In particular, each client trains a new local tree based on the prediction obtained from the global model, then it sends this tree to the server, which aggregates them by concatenation in the order of arrival. This process repeats until the desired number of trees (or training rounds) is achieved. Such an approach is implemented in the Flower framework. Finally, the schema in ~\Cref{fig:flower_bagging} summarizes the algorithm.

\begin{figure}[t]
    \centering
    \includegraphics[width=0.7\linewidth]{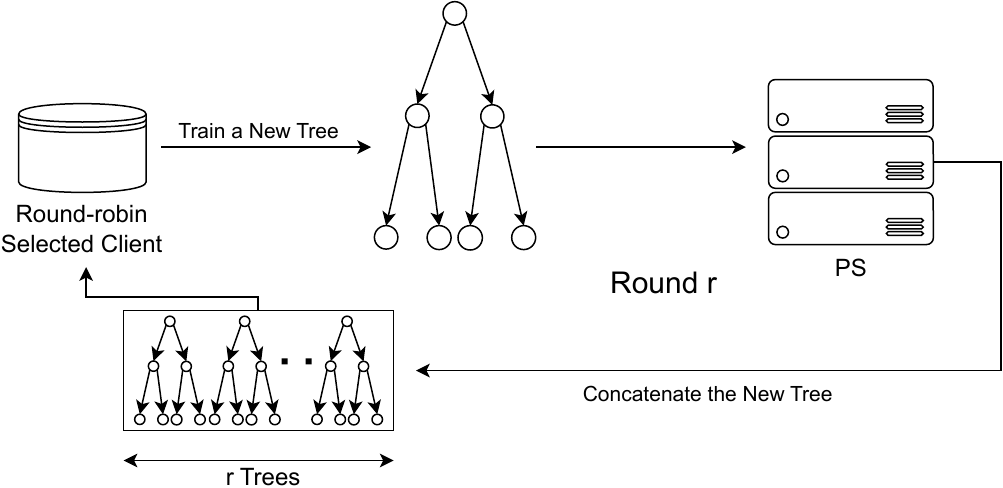}
    \caption{Flower Cyclic algorithm schema.}
    \label{fig:flower_cyclic}
\end{figure}

\mypar{Flower XGBoost Cyclic}\label{sub:xgboost_cyclic}
In this protocol, depicted in \Cref{fig:flower_cyclic}, the global tree ensemble is sequentially updated by different clients. At each round, the server sends the current ensemble to a selected client in a round-robin manner. The client then trains and concatenates a new tree (or trees for multiclass classification) to the ensemble before returning it to the server. This iterative process continues until the ensemble reaches a predefined number of trees.
\begin{figure}[t]
    \centering
    \includegraphics[width=0.8\linewidth]{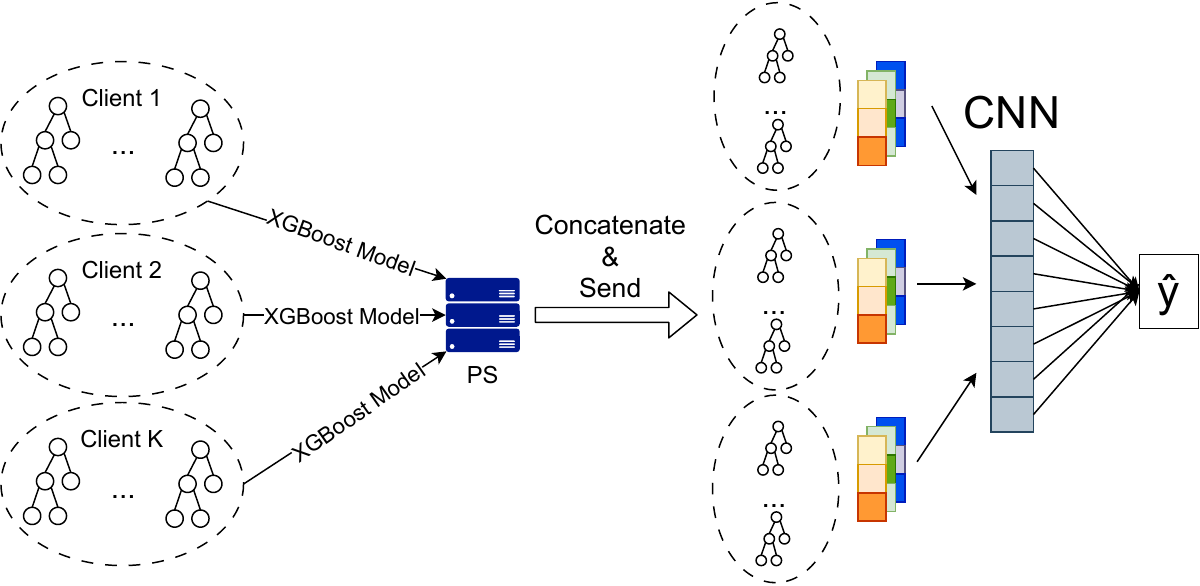}
    \caption{FedXGBllr algorithm schema.}
    \label{fig:fedxgbllr}
\end{figure}

\mypar{FedXGBllr~\cite{ma_gradient-less_2023}}
\label{sub:fedxgbllr}
As shown in \Cref{fig:fedxgbllr}, this protocol introduces a two-phase training strategy to federate XGBoost. In the first phase (first round), clients independently train \acrshort{XGBoost} models and send them to the server, which aggregates (concatenation) them into a global ensemble. The second phase (all subsequent rounds) involves constructing a dataset where each sample is represented by the leaf values it reaches across all trees in the aggregated ensemble. This dataset is then used as input to a federated one-dimensional \acrlong{CNN} (\acrshort{CNN} 1-D), which is trained to predict the sample labels. The intuition is that the learned parameters of the CNN correspond to the learning rates assigned to each tree.

\begin{figure}[t]
    \centering
    \includegraphics[width=0.58\linewidth]{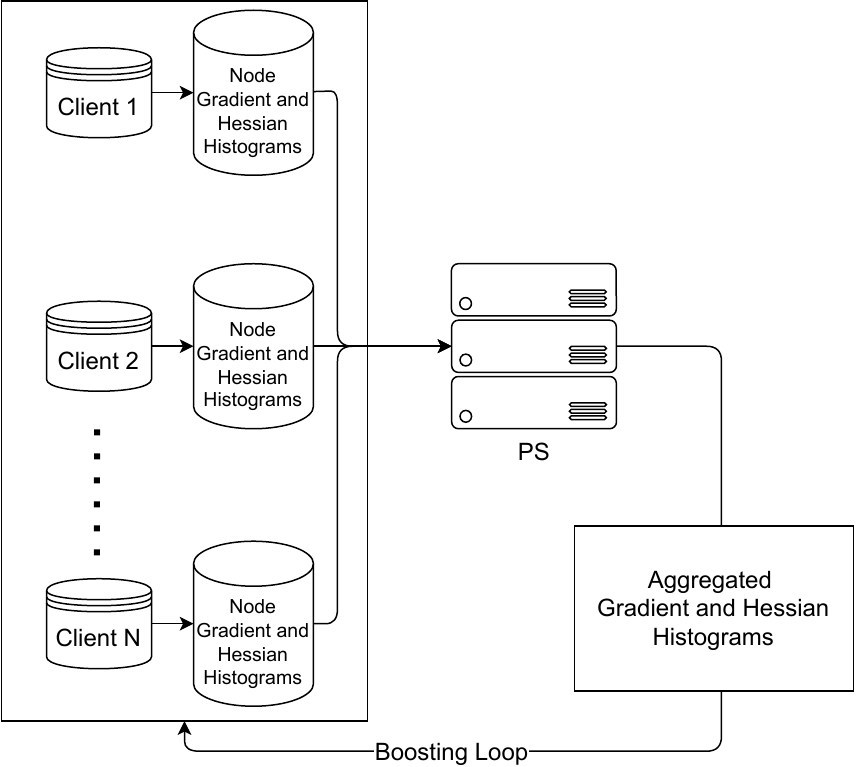}
    \caption{NVFlare Histogram-based algorithm schema.}
    \label{fig:nvflare_histogram_aggregation}
\end{figure}
\mypar{NVFlare - XGBoost Histogram Aggregation}
\label{sub:nvflare_histogram_aggregation}
Developed by NVIDIA, NVFLARE supports federated XGBoost training~\cite{roth_nvidia_2022} by directly federating the XGBoost library. Among the implemented variants, NVFLARE enables histogram aggregation of gradient and Hessian statistics. Unlike the previously described mechanisms, a histogram-based federated algorithm trains each single tree in a distributed manner. At each training round, clients compute and send histograms with gradient and Hessian values to the server, as depicted in \Cref{fig:nvflare_histogram_aggregation}. The server aggregates the histograms by summing them across all clients and returns the resulting global histograms. This global view enables each client to determine the best next split in the tree based on the combined data distribution. This is possible due to the additive properties of gradients and Hessians. Specifically, as discussed in \Cref{sub:xgboostgdbt}, each tree node maintains an aggregated gradient $G$ and Hessian $H$, computed by summing the individual gradients and Hessians of all samples that fall into that node. Therefore, when histograms are aggregated across clients, it is equivalent to increasing the number of samples contributing to the statistics of each node.

\mypar{FedTree~\cite{li_fedtree_2023}}\label{sub:fedtree}
FedTree provides a custom \acrshort{GBDT} implementation. Like NVFlare, it employs histogram aggregation to train tree ensembles in a privacy-preserving manner. However, the key distinction is that in FedTree, the server does not send the aggregated histograms back to the clients. Instead, it uses the aggregated histograms to compute the node information, which is then sent to the clients. This information also includes the aggregated gradient $G$ and Hessian $H$ values. In addition, the authors state that the framework offers three levels of protection. The first level, $L_0$, provides no protection. The second level, $L_1$, protects local histograms by using \textit{secure aggregation}~\cite{bonawitz_practical_2016} for horizontal \acrshort{FL} and \acrshort{HE} for vertical \acrshort{FL}. Finally, the highest level, $L_2$, applies \acrshort{DP} to prevent information leakage from histograms at both the server and client sides.
\section{Related Work} \label{sec:relwork}

The research community has put significant effort into studying adversarial \acrshort{ML}. With the spread of \acrshort{FL}, new threats have emerged~\cite{enthoven_overview_2020,liu_threats_2022,kumar_impact_2023}. Specifically, \acrshort{FL} systems can be vulnerable to several types of attacks, including free-rider attacks~\cite{fraboni_free-rider_2021}, in which participants do not contribute to the training process but still benefit from the global model; utility attacks~\cite{bagdasaryan_how_2019,sun_can_2019}, where adversaries manipulate the training process to gain an advantage; and privacy attacks~\cite{rigaki_survey_2024}, where sensitive information about the underlying training dataset is at risk. In this work, we focus on privacy attacks and their countermeasures. Therefore, the following describes the main related privacy attacks and defenses.

\mypar{Attacks} Several prior works focused on privacy attacks in \acrshort{FL}, specifically targeting federated \acrfullpl{ANN}. Plain gradient sharing has been proven to leak sensitive information about the training data~\cite{zhu_deep_2019,melis_exploiting_2019}. Furthermore, different works have improved the effectiveness of gradient inversion attacks in various scenarios. For instance, such attacks can be performed in the presence of gradient compression~\cite{yang_using_2023}, when the adversary does not have access to the model parameters~\cite{zhang_stealing_2023}, or to reduce reconstruction complexity~\cite{xu_agic_2022}. 
More recently, J. C. Zhao et al. introduced Loki~\cite{zhao_loki_2024}, an attack that breaks the anonymity of model aggregation and is effective against secure aggregation. Another type of attack leverages a \acrfull{GAN} to generate synthetic data that mimics the training dataset~\cite{hitaj_deep_2017}. This approach uses the global model as a discriminator to differentiate between real and synthetic data. The attacker then injects synthetic samples into the training process, gradually refining the \acrshort{GAN} to produce increasingly realistic data. Z. Wang et al.~\cite{wang_beyond_2018} propose an improvement to the \acrshort{GAN} attack, where the adversary is also able to extract user-specific private information without interfering with the training process. A further improvement is presented in GRNN~\cite{ren_grnn_2022}, where the attacker can recover private information from shared gradients without the need for class labels. In the same scenario, H. Wu et al. introduce a class-property inference attack~\cite{wu_federated_2022}, where the adversary aims to infer properties of a specific class in the dataset, focusing on tabular data.

In the tabular data domain, the authors of TabLeak~\cite{vero_tableak_2023} introduce a robust evaluation method for data reconstruction, which we leverage in our study, as it specifically targets tabular datasets.

While most privacy \acrshort{FL} attacks have been studied in the context of \acrshort{ANN}, much less attention has been given to tree-based models. Some works have explored privacy attacks in vertical \acrshort{FL} for tree-based models~\cite{chen_fia-te_2024,takahashi_eliminating_2023,luo_feature_2021}. However, horizontal settings remain unexplored. Therefore, our work aims to close this gap.

\mypar{Defenses}
Several defense mechanisms have been proposed to enhance privacy in \acrshort{FL} systems. Hardware-based solutions, such as secure enclaves~\cite{law_secure_2020}, provide isolated execution environments that protect computations on the server side.
Techniques such as \acrfull{HE}~\cite{cheng_secureboost_2021,liu_boosting_2020,cong_sortinghat_2022} offer strong theoretical privacy guarantees by enabling model aggregation on encrypted data. In \acrshort{FL}, HE is used to protect client updates from server-side privacy threats. However, its practicality is limited due to the restricted set of operations that can be performed on ciphertexts. Similarly, \acrfull{MPC}~\cite{lindell_secure_2005,wu_privacy_2020} allows multiple parties to jointly compute functions without revealing their inputs. In \acrshort{FL}, the most widely adopted class of \acrshort{MPC} is \textit{secure aggregation}~\cite{bonawitz_practical_2016}, which targets server-side threats. Nonetheless, these cryptographic approaches often suffer from substantial computational and communication overhead~\cite{li_practical_2019,acar_survey_2018}. Alternative strategies—such as \acrfull{DP}~\cite{li_fedtree_2023} and \acrfull{LSH}~\cite{li_practical_2019}—seek to protect sensitive information by reducing the granularity of shared data. For instance, \acrshort{DP} injects noise into model updates to protect against privacy threats on both the client and server sides. Despite their promise, these non-cryptographic defenses introduce trade-offs between privacy and utility, as the noise added for protection can degrade model performance~\cite{li_trade-off_2021}. Among the previously discussed approaches, those currently implemented in \acrshort{FL} tree-based systems include \acrlong{DP}, \acrlong{MPC} (via secure aggregation), and \acrlong{HE}. \acrshort{DP} protects against both client- and server-side privacy threats, whereas \acrshort{MPC} and \acrshort{HE} offer protection only against server-side threats. Regarding frameworks, NVFLARE supports secure aggregation, while FedTree implements all three approaches.

\subsection{Research Gap and Motivation} \label{sec:motivation}
\acrlong{FL} has been extensively studied in the context of \acrfullpl{ANN}, with several works proposing attacks and corresponding countermeasures to address privacy concerns. However, to the best of our knowledge, no prior work has demonstrated a dataset reconstruction attack targeting horizontal federated tree-based systems. This lack reveals a critical gap in the security analysis of federated gradient boosting models, despite their growing use in privacy-sensitive applications.

Tree-based \acrshort{FL} frameworks differ substantially from their \acrshort{ANN}-based counterparts in terms of model structure, aggregation mechanisms, and information leakage patterns. While adversarial research on \acrshortpl{ANN} has inspired the development of privacy-preserving techniques, tree-based \acrshort{FL} systems remain largely untested against realistic adversarial scenarios. To address this gap, we propose a novel dataset reconstruction attack that exploits the inherent properties of boosted \acrshortpl{DT} and how they are federated in current \acrshort{FL} frameworks.

Importantly, our threat model—discussed in \Cref{sec:threatmodel}—assumes that the attacker controls a client. Consequently, the attack can be executed by exploiting only client-side information. As previously explained, privacy defenses such as \acrlong{HE} and \acrlong{MPC} are designed to prevent server-side privacy leakage and, therefore, do not alter the information received by clients. As a result, they are ineffective in our threat model. Moreover, applicable defenses such as \acrlong{DP}—specifically in its \acrlong{LDP}~\cite{li_trade-off_2021} form—have not yet been evaluated against real-world attacks in horizontal \acrshort{FL} scenarios. To address this gap, we evaluate the effectiveness of \acrshort{DP} against \name.

In summary, by introducing a novel attack, we aim to systematically uncover key vulnerabilities in tree-based \acrshort{FL} systems that allow adversaries to reconstruct private training datasets with measurable accuracy and to evaluate whether existing defenses can mitigate such attacks. This work is further motivated by the goal of proposing theoretical modifications to current frameworks, ultimately paving the way for more robust and privacy-preserving gradient boosting \acrshort{FL} architectures.

\section{Threat Model}\label{sec:threatmodel}
This section defines the threat model by defining the adversary's capabilities, objectives, and target.

\mypar{Adversary's Capabilities}
We assume the adversary can control a client involved in a horizontal \acrshort{FL} system where the server is a \textit{trusted entity}. Consequently, they have full white-box access to the trained global model, allowing them to inspect its internal structure. In particular, they can access all the information the clients have at training time. Such information depends on the considered implementation but, in general, includes the trained \acrshortpl{DT}, their splits, leaf values, hyperparameters, and the aggregated gradient and Hessian values used during training (all definitions about tree-related concepts in \Cref{sub:xgboostgdbt}).
The adversary is modeled as an honest-but-curious participant, as in prior work~\cite{vero_tableak_2023}, meaning they strictly follow the protocol without attempting to disrupt the system. However, they seek to infer sensitive information about other participants’ training data by analyzing model parameters and training statistics available on the client side. Unlike prior works, which assume the honest-but-curious adversary is located on the server, we assume it resides on one of the clients. This scenario is particularly relevant in \acrshort{FL} settings where participants can be market competitors and training data is a valuable asset. Moreover, because the adversary behaves according to the protocol, they are harder to detect—the attack can be conducted entirely ``offline.''

\mypar{Adversary's Objective} The adversary aims to reconstruct the training dataset from other participants in the \acrshort{FL} process by analyzing the model parameters and training statistics exchanged during training. The goal is to infer sensitive information about the training data, such as the presence of specific samples or the distribution of features and labels in the training data. This information could be used to gain a competitive advantage or to compromise the privacy of the participants.

\mypar{Targeted Implementations} The adversary targets other clients' datasets by exploiting vulnerabilities in 5 main implementations of federated tree-based systems, including Flower Bagging and Cyclic~\cite{beutel_flower_2022}, NVFlare histogram-based~\cite{roth_nvidia_2022}, FedXGBllr~\cite{ma_gradient-less_2023}, and FedTree~\cite{li_fedtree_2023}. These implementations represent diverse approaches, each varying in the amount of information exposed during training. We consider them representative of the \acrshort{SOTA} in federated tree-based models and relevant for practical deployments.
\section{\name: Dataset Reconstruction Attack}
\label{sec:approach}

\begin{figure}[t]
    \centering
    \includegraphics[width=0.9\linewidth]{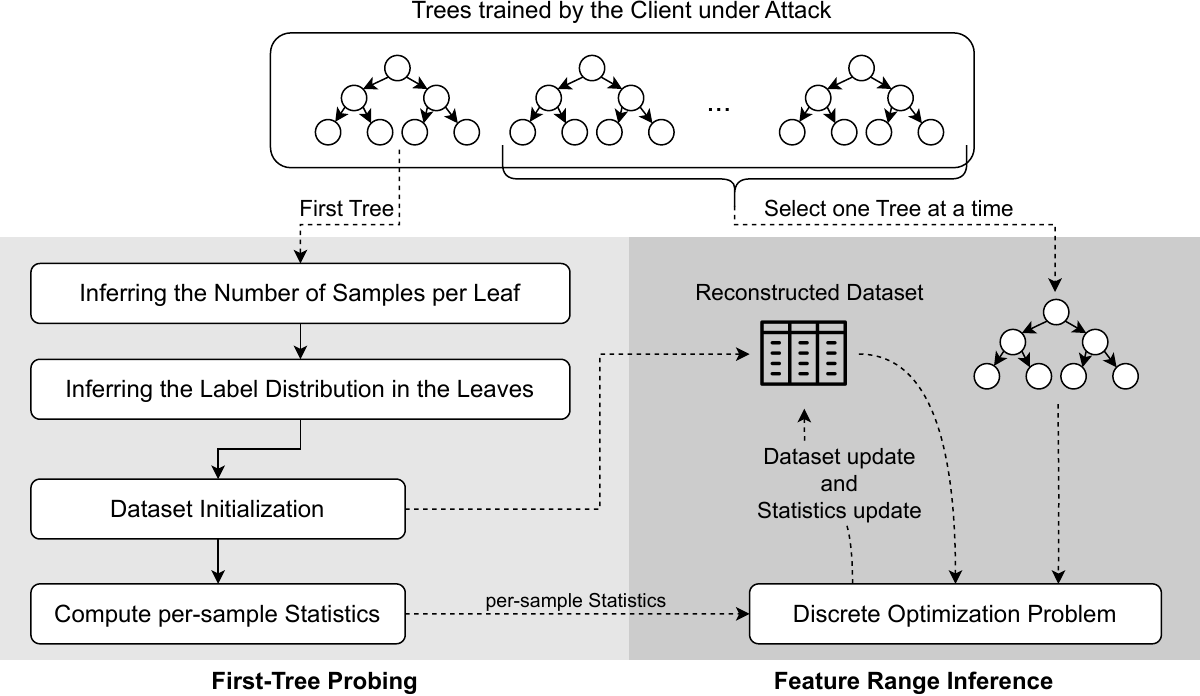}
    \caption{\name attack schema.}
    \label{fig:attack_schema}
\end{figure}

We propose a novel dataset reconstruction attack, called \name, to assess the privacy risks associated with training federated tree-based models. \name specifically targets the implementations of \acrshort{XGBoost} and \acrshort{GBDT} in horizontal \acrshort{FL} settings.

The attack reconstructs clients' training data by exploiting information accessible to an honest-but-curious adversary. As defined in \Cref{sec:threatmodel}, our threat model—guided by existing \acrshort{FL} implementations—assumes that the adversary, by controlling a client within the system, can access the trained model(s) in plaintext. Consequently, the adversary can inspect tree structures, aggregated gradients, and Hessians. Additionally, they can observe model parameters such as the base score and learning rate, which remain accessible during training. Since these statistics and parameters are fundamental to understanding the internals of \name, we provide all relevant definitions related to gradient boosting in \Cref{sub:xgboostgdbt}.

The intuition behind our approach is that an attacker, by controlling a client, can exploit the sequential nature of boosted trees. By analyzing the decision splits and gradient statistics shared by other clients, the attacker can infer information about the training data that influenced the tree construction. The attack can be performed at each round of training. However, the more rounds that have been completed, the more trees have been trained. Therefore, the attack will be more accurate. 
As depicted in Figure~\ref{fig:attack_schema}, \name can be divided into two main phases that need to be repeated for each client the adversary targets. The first phase, \textit{First-Tree Probing}, examines the first tree trained by the client under attack to infer sample counts and label distributions, and to initialize the reconstructed dataset based on the observed tree splits. The second phase, \textit{Feature Range Inference}, refines the reconstruction by solving an optimization problem on each subsequent tree trained by the client under attack.
Following, we provide a detailed explanation of the attack by describing each phase in depth. Finally, we discuss its adaptation to different federated gradient boosting implementations. Note that the attack formalization pertains to a \textit{binary classification} task. However, it can be extended to \textit{multi-class classification}, for which we provide a formalization in \Cref{appendix:multiclass}.

\subsection{First-Tree Probing}\label{sub:fist-tree-prob}

As shown in \Cref{sub:fltrees_background}, regardless of the specific federated implementation, the client receives the global model from the \acrfull{PS} at least once. This means the controlled client has visibility into the global model, which contains trees contributed by all other clients. In this phase, the attacker targets a specific client and selects the \textit{first tree} trained by that client. The motivation for selecting the first tree is that it is built on predictions obtained using the base score. As we will explain, this plays a crucial role in \name. As depicted in Figure~\ref{fig:attack_schema}, this first phase of \name can be further divided into four steps, aimed at inferring the number of samples per leaf, inferring the label distribution in the leaves, initializing the dataset, and computing per-sample statistics.

\mypar{Inferring the Number of Samples per Leaf} After clients receive the aggregated trees from the \acrshort{PS}, the adversary can infer the number of samples assigned to each leaf $j$ by analyzing the first tree trained by the chosen victim.
Before discussing how we can obtain this information, let us first analyze the gradient and Hessian statistics for a binary classification task. In particular, given a log loss (or binary cross-entropy loss) function, these statistics can be obtained~\cite{chen15, Goodfellow-et-al-2016} for each sample $i$, starting from \Cref{eq:hessian_gradient}:
\begin{align}
    g_i = p_i - y_i,  \quad   h_i = p_i \cdot (1 - p_i),
    \label{eq:hessian_gradient_probability}
\end{align}
where the probability score \(p_i = \sigma(x)\), $x$ is the sum of outputs of the previously trained trees, \( \sigma(x) = \frac{1}{1 + e^{-x}} \), and \( y_i \) is the label of the \( i \)-th sample. Now, by analyzing the first tree, we know that $p_i$ only depends on the base score, which is the constant global bias of our model. Considering base score is already a probability, for each sample during the training of the first tree, we have $p_i = \sigma(\sigma^{-1}(base\_score)) = base\_score$. We can, therefore, write:
\begin{equation}
\small
    h_i = base\_score \cdot (1 - base\_score),
\end{equation}
and the total aggregated Hessian for a given leaf $j$ as:
\begin{equation}
\small
    H_j = \sum_{i=1}^{N_j} h_{ij} = N_j \cdot base\_score \cdot (1 - base\_score),
\end{equation}
where \( N_j \) represents the number of samples assigned to the leaf $j$. Solving for \( N_j \), we obtain:
\begin{equation}
\small
    N_j = \frac{H_j}{base\_score \cdot (1 - base\_score)},
\end{equation}
By disposing of both the base score and the total aggregated Hessian for each leaf in the tree, we can solve the above equation.

\mypar{Inferring the Label Distribution in the Leaves} Once \name infers the number of samples per leaf, it can infer the \textit{distribution of labels} within each leaf, forming the foundation for \textit{dataset initialization}. Indeed, by knowing the label distribution and the number of samples, an adversary can initialize a dataset with the same size and distribution as the targeted one. To obtain this information, we can exploit the aggregated gradient for each leaf $j$. In gradient boosting, this gradient is given by the formula:
\begin{equation}
\small
    G_j = - \frac{leaf\_value_j}{\eta} \cdot (H_j + \lambda),
\end{equation}
where $\eta$ is the learning rate and $\lambda$ is the regularization parameter. Since the gradient for each sample depends on its label, we leverage the following expression to differentiate between samples labeled as 0 and 1 (binary classification):
\begin{equation}
\small
    G_j = \sum_{i=1}^{N_j} g_{ij} = N^{(0)}_{j} \cdot base\_score - N^{(1)}_{j} \cdot (1 - base\_score),
\end{equation}
where \( N^{(0)}_{j} \) and \( N^{(1)}_{j} \) denote the number of samples with labels 0 and 1 in the leaf $j$. Given that the total number of samples in the leaf satisfies \( N^{(0)}_{j} + N^{(1)}_{j} = N_j \), we can write:
\begin{equation}
\small
    N^{(1)}_{j} = N_j \cdot base\_score - G_j, \quad N^{(0)}_{j} = N_j - N^{(1)}_{j},
\end{equation}
and, disposing of both $G$ for each leaf and the base score, the adversary can solve the above equations to finally compute the label distribution within a certain leaf.

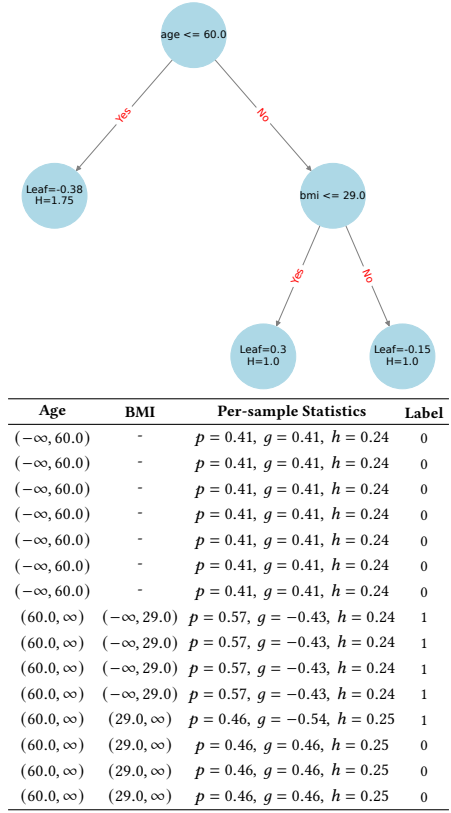
\begin{figure}[t]
    \centering
    \begin{minipage}[c]{\linewidth}
        \centering
        \includegraphics[width=0.65\linewidth]{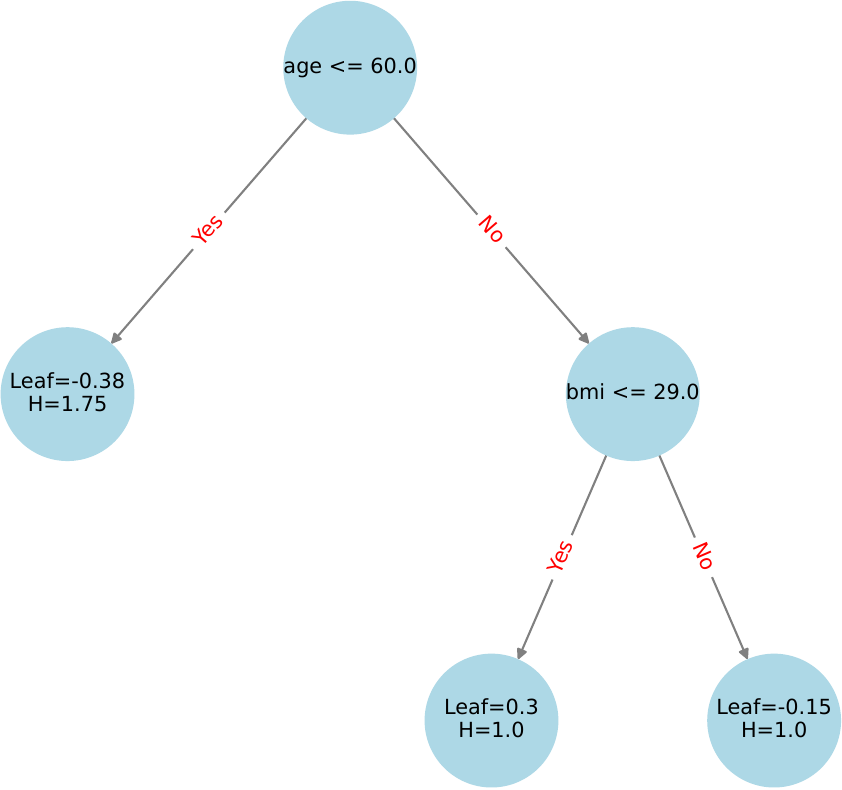}
    \end{minipage}\\
    \begin{minipage}[c]{\linewidth}
        \centering
        \begin{tikzpicture}
        \scriptsize
        \matrix (m) [matrix of nodes,
                     row sep=-\pgflinewidth, 
                     column sep=-\pgflinewidth,
                     nodes={draw=none, anchor=center}] {
                \hline
                \textbf{Age} & \textbf{BMI} & \textbf{Per-sample Statistics} & \textbf{Label} \\
                \hline
                $(-\infty, 60.0)$ & - & $p=0.41,\ g=0.41,\ h=0.24$ & 0 \\
                $(-\infty, 60.0)$ & - & $p=0.41,\ g=0.41,\ h=0.24$ & 0 \\
                $(-\infty, 60.0)$ & - & $p=0.41,\ g=0.41,\ h=0.24$ & 0 \\
                $(-\infty, 60.0)$ & - & $p=0.41,\ g=0.41,\ h=0.24$ & 0 \\
                $(-\infty, 60.0)$ & - & $p=0.41,\ g=0.41,\ h=0.24$ & 0 \\
                $(-\infty, 60.0)$ & - & $p=0.41,\ g=0.41,\ h=0.24$ & 0 \\
                $(-\infty, 60.0)$ & - & $p=0.41,\ g=0.41,\ h=0.24$ & 0 \\
                $(60.0, \infty)$ & $(-\infty,29.0)$ & $p=0.57,\ g=-0.43,\ h=0.24$ & 1 \\
                $(60.0, \infty)$ & $(-\infty,29.0)$ & $p=0.57,\ g=-0.43,\ h=0.24$ & 1 \\
                $(60.0, \infty)$ & $(-\infty,29.0)$ & $p=0.57,\ g=-0.43,\ h=0.24$ & 1 \\
                $(60.0, \infty)$ & $(-\infty,29.0)$ & $p=0.57,\ g=-0.43,\ h=0.24$ & 1 \\
                $(60.0, \infty)$ & $(29.0,\infty)$ & $p=0.46,\ g=-0.54,\ h=0.25$ & 1 \\
                $(60.0, \infty)$ & $(29.0,\infty)$ & $p=0.46,\ g=0.46,\ h=0.25$ & 0 \\
                $(60.0, \infty)$ & $(29.0,\infty)$ & $p=0.46,\ g=0.46,\ h=0.25$ & 0 \\
                $(60.0, \infty)$ & $(29.0,\infty)$ & $p=0.46,\ g=0.46,\ h=0.25$ & 0 \\
                \hline \\
            };
        
        \end{tikzpicture}
    \end{minipage}
    \caption{\acrfull{DT} with its reconstructed dataset initialized after the First-Tree Probing phase. In this example, we use $\lambda = 1$, $learning\_rate = 0.3$, and $base\_score = 0.5$.}
    \label{fig:attack_example}
\end{figure}

\mypar{Dataset Initialization} At this stage, the adversary knows both the number of samples per leaf and their label distributions. Therefore, they can generate the exact number of samples assigned from the training set to that leaf, with the exact label distribution. Moreover, in tree-based models, each sample reaches a leaf by following a path dictated by its feature values. \name exploits this property to impose constraints on the range of features in a sample through the observed leaf assignments. In summary, at this step, the adversary can generate the first instance of the reconstructed dataset (an example in \Cref{fig:attack_example}).

\mypar{Compute per-sample Statistics} This final step is essential for executing the second phase and aims to complete each reconstructed sample with three statistics used during training by gradient boosting algorithms.
In particular, for each generated sample \( i \), we compute the gradient \( g_i \) and the Hessian \( h_i \) according to \Cref{eq:hessian_gradient_probability}, using the probability score \( p_i \) computed as:  
\begin{align}
\small
    \label{eq:attack_gradient}
    p_i &= \sigma(\sigma^{-1}(base\_score) + \sum_{t=1}^{T} leaf\_score_i^{(t)}).
\end{align}
where \( \sigma^{-1}(x) = \log\left(\frac{x}{1 - x}\right) \), \( T \) is the number of already analyzed trees (in the first phase $T=1$), and \(leaf\_score_i^{(t)}\) is the prediction value assigned to the leaf of the tree \( t \) in which the \( i \)-th sample falls.

\subsection{Feature Range Inference}
Once the initial feature ranges have been inferred, \name refines the generated dataset by formulating a \acrfull{MILP} problem to determine the precise leaf assignment for each sample in each subsequent tree. In other words, after analyzing the first tree, \name analyzes and solves a \acrshort{MILP} problem for each other tree trained by the client under attack. Since each sample follows a unique path through the tree, based on its feature values, we leverage the feature constraints accumulated so far to restrict the possible set of leaves a sample can reach. This defines a subset of all possible leaves, which we use as constraints in our optimization problem, which minimizes the discrepancy between reconstructed and original aggregated gradients and Hessians, progressively refining feature estimates.

Looking at a tree in the ensemble, for each sample, we define a binary assignment variable \( x_{ij} \), which indicates whether sample \( i \) is assigned to leaf \( j \). The constraints of the optimization problem enforce that each sample is assigned to exactly one leaf and that this leaf belongs to the set of reachable leaves based on previously inferred feature ranges. As for any optimization problem, we need to define the \textit{constants}, the \textit{variables}, the \textit{constraints}, and the \textit{objective function}. Following the formalization of the designed problem.

\begin{definition}
\small
\mypar{Constants} We define the following constants:
\begin{itemize} 
    \item \( I = \{1, \ldots, n\} \): set of training samples.
    \item \( J = \{1, \ldots, m\} \): set of leaves.
    \item \( G_j \): aggregated gradient of leaf \( j \).
    \item \( H_j \): aggregated Hessian of leaf \( j \).
    \item \( p_i \): probability score of sample \( i \) from previous trees.
    \item \( g_i \): gradient of sample \( i \), computed from \( p_i \) and its label.
    \item \( h_i \): Hessian of sample \( i \), computed from \( p_i \) and its label.
    \item \( L_i \subseteq J \): set of leaves that sample \( i \) can reach, based on inferred feature constraints.
\end{itemize}

\mypar{Variables} We define the following variables:
\begin{itemize}
    \item \( x_{ij} \in \{0,1\} \): binary variable indicating whether sample \( i \) is assigned to leaf \( j \).
\end{itemize}

\mypar{Constraints}
Each sample must be assigned to exactly one leaf:
\begin{equation}
    \sum_{j \in J} x_{ij} = 1, \quad \forall i \in I.
\end{equation}

The leaf assignments must respect the inferred feature constraints:
\begin{equation}
    x_{ij} = 0, \quad \forall i \in I, \forall j \notin L_i.
\end{equation}

\mypar{Objective Function} The objective is to minimize the discrepancy between the reconstructed and original gradients and Hessians:
\begin{equation}
    \min \sum_{j \in J} (\sum_{i \in I} x_{ij} \cdot g_i - G_j)^2 + (\sum_{i \in I} x_{ij} \cdot h_i - H_j)^2.
\end{equation}

\end{definition}

By iteratively solving the above optimization problem, we update the feature ranges for each sample and the gradient and Hessian values by re-executing the last step of the previous phase (Compute per-sample Statistics). Following this process, we progressively refine the estimated feature values of each sample. At the end of the process, \name achieves a reconstruction of the training data that is consistent with the observed model.

\subsection{\name Details} \label{sub:details_approach}

We now present how our attack can be adapted to target each implementation discussed in \Cref{sub:fltrees_background} (Flower Bagging and Cyclic~\cite{beutel_flower_2022}, NVFlare histogram-based~\cite{roth_nvidia_2022}, FedXGBllr~\cite{ma_gradient-less_2023}, and FedTree~\cite{li_fedtree_2023}). In addition, we discuss how these implementations impact the reconstruction granularity. Specifically, while Flower Bagging, Cyclic, and FedXGBllr allow for targeting and reconstructing \textit{local} clients' datasets, histogram-based systems such as NVFlare and FedTree only enable reconstruction of the aggregated \textit{global} dataset—i.e., a union of all clients' datasets. We adapt \name accordingly, detailing the differences in the attack implementation for each case.

\subsubsection{Flower \acrshort{XGBoost} Bagging and Cyclic}
In these two implementations, the first tree in the model is not necessarily trained by the client under attack. Indeed, the \acrfull{PS} concatenates the trees in arrival order. Furthermore, in both implementations, each client trains local trees using models from previous rounds, which incorporate trees trained by other clients. As a result, the aggregated gradient and Hessian statistics at each round are computed on samples that do not necessarily belong to the client under attack. Therefore, at each round, we infer feature ranges for each sample and perform a weighted average of the leaf values from trees not trained by the client under attack, using the Hessian values as weights. Consequently, we approximate the true leaf in which the sample is likely to fall across the trees of other clients while preserving the feature range constraints that are relevant only to trees trained by the victim client. Finally, looking only at the trees trained by the victim client, we proceed as usual by solving the optimization problem. 

\begin{figure}[t]
    \centering
    \includegraphics[width=0.66\linewidth]{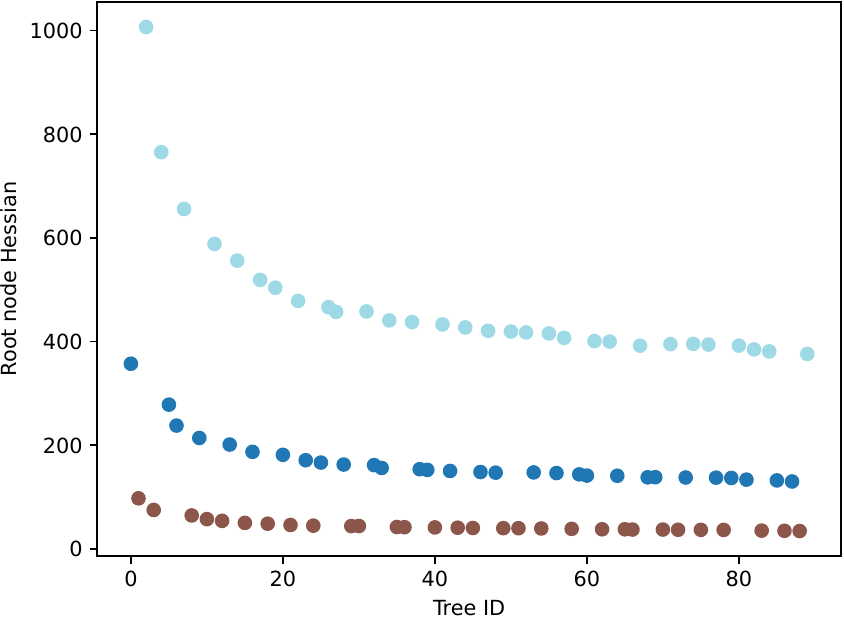}
    \caption{Root node Hessian for 90 trees trained using the Flower \acrshort{XGBoost} Bagging implementation. Each client trains 30 trees, which are then identified under the same client by minimizing the distance with the trees in the previous round.}
    \label{fig:tree_clustering}
\end{figure}

Flower \acrshort{XGBoost} Bagging differs from the Cyclic variant in that client ordering is not fixed; instead, it depends on client arrival times or follows a randomized schedule. Therefore, to recognize the trees from the same client, we look at the Hessian and gradient statistics. In particular, the idea is that, given the data on which a single client trains a tree are always the same across rounds, it is possible to find a relation between a tree at round $r$ and a tree trained by the same client at round $r-1$ by searching for the minimum distance between the hessian values in a tree at round $r$ and a tree at round $r-1$. Still, this mechanism only affects the second phase. In case of issues in reconstructing the chain of trees trained by the same client, the adversary can still use the \textit{First-Tree Probing} phase alone using only the first received trees. In~\Cref{fig:tree_clustering} we plot the Hessian values of the root nodes for an example using 3 clients.
Moreover, an honest-but-curious adversary lacks direct knowledge of the mapping between tree ordering and client identities, but while it is not possible to attribute reconstructed samples to specific clients, the adversary can still infer that certain samples originate from the same client.

\subsubsection{FedXGBllr} For such an implementation, the attack doesn't need to be adapted. We extract the victim client's local trees and apply the method described in Section~\ref{sec:approach}. The first round of the protocol, where the trees contained in the \acrshort{XGBoost} models are shared, is enough to reconstruct the entire training set. The attack enables user-level reconstruction, as the models are trained separately on clients' data, and the locally trained trees are shared with all other clients (including the adversary) in the initialization phase.

\subsubsection{NVFlare and FedTree} These two implementations aggregate histograms to determine, at each round, the best splits that each client should perform during the training of a single tree. As a result, each client receives only aggregated gradient and Hessian statistics derived from the combined histograms (i.e., the sum of the clients' histograms). The attack proceeds as described in the general formalization.  However, in this case, the adversary can only reconstruct the \textit{global} dataset (i.e., the union of all clients’ datasets), since these systems expose only aggregated model updates rather than individual local trees. Consequently, the attacker observes the boosting of \textit{global} trees.
\section{Experimental Evaluation}
Our experimental evaluation aims to test \name to address the following research questions:

\noindent\textbf{RQ1}: Is our attack effective 
across all considered tree-based \acrshort{FL} systems? Does the \textit{Feature Range Inference} phase enhance effectiveness, and what is the impact of the \textit{First-Tree Probing} phase alone?

\noindent\textbf{RQ2}: How does tree depth influence the effectiveness of the attack in the considered systems? Specifically, does increased depth and potential overfitting help the dataset reconstruction?
  
\noindent\textbf{RQ3}: Can the attack's effectiveness be mitigated by applying classical defenses such as \acrshort{DP} during training?

\begin{table*}[t]
\centering
\scriptsize
\caption{F1-score, and \acrfull{AUC} on the test set for the binary classification task on the Stroke Prediction and Pima Indians Diabetes datasets.}
\label{tab:performance}
\setlength{\tabcolsep}{4pt}
\renewcommand{\arraystretch}{1.2}
\begin{tabular}{|c|c|cc|cc|cc|cc|cc|cc|}
\hline
\multirow{2}{*}{Dataset} & \multirow{2}{*}{Implementation} & \multicolumn{2}{c|}{Depth 3} & \multicolumn{2}{c|}{Depth 4} & \multicolumn{2}{c|}{Depth 5} & \multicolumn{2}{c|}{Depth 6} & \multicolumn{2}{c|}{Depth 7} & \multicolumn{2}{c|}{Depth 8} \\
 &  & F1 & AUC & F1 & AUC & F1 & AUC & F1 & AUC & F1 & AUC & F1 & AUC \\
\hline
\multirow{5}{*}{Stroke}
 & FedXGBllr
 & 0.215 & 0.635
 & 0.218 & 0.628
 & 0.183 & 0.620
 & 0.157 & 0.617
 & 0.105 & 0.608
 & 0.097 & 0.614
 \\
 & Flower XGBoost Bagging
 & 0.213 & 0.755
 & 0.215 & 0.750
 & 0.185 & 0.750
 & 0.176 & 0.746
 & 0.138 & 0.731
 & 0.160 & 0.751
 \\
 & Flower XGBoost Cyclic
 & 0.237 & 0.748
 & 0.157 & 0.738
 & 0.158 & 0.731
 & 0.162 & 0.760
 & 0.176 & 0.734
 & 0.158 & 0.709
 \\
 & NVFlare
 & 0.165 & 0.767
 & 0.203 & 0.774
 & 0.203 & 0.767
 & 0.164 & 0.775
 & 0.103 & 0.744
 & 0.128 & 0.761
 \\
 & FedTree
 & 0.207 & 0.766
 & 0.200 & 0.764
 & 0.209 & 0.766
 & 0.110 & 0.767
 & 0.130 & 0.768
 & 0.108 & 0.769
 \\
\hline
\multirow{5}{*}{Pima}
 & FedXGBllr
 & 0.587 & 0.816
 & 0.603 & 0.806
 & 0.572 & 0.810
 & 0.531 & 0.806
 & 0.607 & 0.809
 & 0.582 & 0.814
 \\
 & Flower XGBoost Bagging
 & 0.638 & 0.769
 & 0.587 & 0.743
 & 0.577 & 0.716
 & 0.506 & 0.689
 & 0.537 & 0.703
 & 0.585 & 0.718
 \\
 & Flower XGBoost Cyclic
 & 0.573 & 0.753
 & 0.506 & 0.669
 & 0.490 & 0.669
 & 0.591 & 0.755
 & 0.549 & 0.719
 & 0.529 & 0.679
 \\
 & NVFlare
 & 0.643 & 0.817
 & 0.676 & 0.834
 & 0.639 & 0.821
 & 0.636 & 0.831
 & 0.627 & 0.832
 & 0.660 & 0.825
 \\
 & FedTree
 & 0.586 & 0.810
 & 0.596 & 0.826
 & 0.606 & 0.798
 & 0.604 & 0.805
 & 0.604 & 0.806
 & 0.610 & 0.815
 \\
\hline
\end{tabular}
\label{tab:model_performance}
\end{table*}

\subsection{Experimental Setup}\label{sub:exp_setup}

We evaluate \name on five different federated systems—four based on \acrshort{XGBoost} and one on \acrshort{GBDT}. In particular, we adapt and evaluate our attack on XGBoost Bagging, Cyclic, and FedXGBllr~\cite{ma_gradient-less_2023}, all implemented in Flower~\cite{beutel_flower_2022}. Additionally, we apply \name to histogram-based implementations in NVFlare~\cite{roth_nvidia_2022} and FedTree~\cite{li_fedtree_2023}. We consider a general scenario with 3 clients collaboratively training a federated gradient boosting model composed of 100 trees with a depth $d$ from 3 to 8 (we perform 6 runs for each implementation by increasing $d$). To evaluate the scalability of \name, in~\Cref{subsub:scalability} we vary the number of clients up to 30. One client acts as an honest-but-curious adversary attempting to extract sensitive information from the other clients performing the \name attack. In the Flower implementations, where \name can target a specific client, we designate one of the two remaining clients as the victim. In histogram-based implementations, since the attack can only reconstruct the global dataset (i.e., the union of all clients' datasets), we compare the reconstructed dataset with the original dataset, which consists of the combined datasets from all clients. From now on, we will refer to the reconstruction targeting a specific client as \textit{local reconstruction}, while the reconstruction of the global dataset, as in the case of histogram-based implementations, will be referred to as \textit{global reconstruction}.
Finally, to limit the computational load of the \acrlong{MILP} problems in our \emph{Feature Range Inference} phase (more details in \Cref{appendix:complexity}), we impose a time constraint of $10$ minutes for each tree analyzed.

\mypar{Hardware and Software Testing Environment} We conduct our experiments on a machine equipped with an Intel 11th Gen Core i7-1165G7 processor (8 cores, 16 threads, 2.80 GHz base clock, 4.70 GHz max turbo) and 16 GB of RAM. The software environment includes Python 3.10, XGBoost 2.1.0, and the latest version of Gurobi~\cite{gurobi}.

\mypar{Datasets} We conduct our experiments using two publicly available binary classification datasets from the healthcare domain, particularly renowned for the privacy of the training data: the \textit{Stroke Prediction Dataset}~\cite{mxfb-sc71-23} and the \textit{Pima Indians Diabetes Dataset}~\cite{choubey2017classification}. The Stroke Prediction Dataset consists of 5110 records, each representing an individual with attributes related to stroke risk factors. It includes 11 features (including the stroke column), categorized into demographic features (\textit{age, gender, marital status, residence type}), medical conditions (\textit{hypertension, heart disease}), lifestyle factors (\textit{smoking status, work type}), and clinical measurements (\textit{average glucose level, BMI}). In summary, it contains 3 numerical and 7 categorical features. This dataset is highly imbalanced, with only 4.87\% of samples corresponding to stroke cases. Therefore, we apply SMOTE-NC~\cite{chawla_smote_2002} to balance the classes. The Pima Indians Diabetes Dataset contains 728 records and 9 features (including the outcome column). The features include \textit{pregnancies}, \textit{OGTT (Oral Glucose Tolerance Test)}, \textit{blood pressure}, \textit{skin thickness}, \textit{insulin}, \textit{BMI}, \textit{age}, and \textit{pedigree diabetes function}. All these features are numerical.

\mypar{Data Distribution} We focus on a scenario where the distribution of data across clients is non-\acrshort{iid}, as it better reflects real-world conditions. As with any other work in the \acrshort{SOTA}, to simulate the non-\acrshort{iid} case, we use the Dirichlet distribution~\cite{sinharay_continuous_2010} ($\alpha=0.3$) to partition the data across the clients.

\begin{figure}[t]
    \centering
    \includegraphics[width=\linewidth]{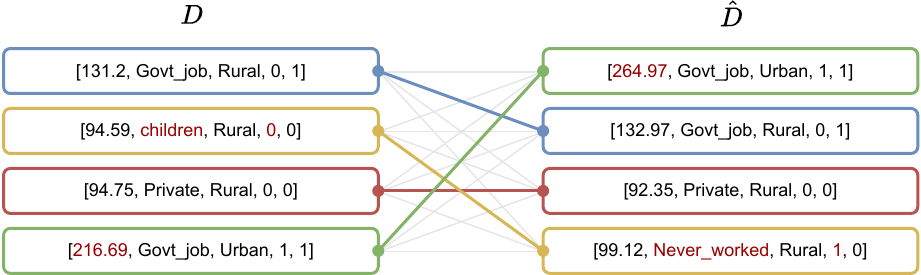}
    \caption{Reconstruction assessment of a 4-sample dataset, using the 5 most important features from Stroke. Each sample in the reconstructed dataset $\hat{D}$ is compared with one in the original dataset $D$. In red, we highlight the wrong reconstructed values. Here, we are able to fully reconstruct the first and third samples (blue and red), the 60\% of the second sample (yellow), and the 80\% of the last sample (green). This leads to an \acrshort{RA} of 85\%.}
    \label{fig:reconstruction_example}
\end{figure}

\begin{table}[t]
    \caption{Tolerance $\epsilon$ for the \acrfull{RA} on the top 5 features of each of the datasets.}
    \centering
    \resizebox{0.9\linewidth}{!}{%
    \begin{tabular}{c|ccccc}
    \toprule
        \textbf{Top 5}&   \textbf{Avg. Glucose} & \textbf{Work}  & \textbf{Residence}  & \textbf{Heart disease} & \textbf{Label} \\ 
         Stroke &  17.03 & cat* & cat* & cat* & cat*\\ \hline
        \textbf{Top 5}&   \textbf{BMI} & \textbf{Diabetes pedigree} & \textbf{Glucose} & \textbf{Age} & \textbf{Label} \\ 
        Pima & 2.50 & 0.105 & 10.06 & 3.77 & cat* \\ \bottomrule
        \multicolumn{6}{c}{\small *categorical, the predicted feature needs to match perfectly with the original.}
    \end{tabular}
    }
    \label{tab:tolerance}
\end{table}

\mypar{Evaluation Metrics}
We evaluate the effectiveness of our approach using three main metrics: the \textit{\acrfull{RA}}~\cite{vero_tableak_2023}, the \textit{F1-score}, and the \textit{\acrfull{AUC}}. They are used to assess the efficacy of the reconstruction attack (\acrlong{RA}) and the performance of the model on the trained task (F1 and \acrshort{AUC}). While the F1-score and the \acrshort{AUC} are well-known classification metrics, the \acrlong{RA} is a metric introduced in TabLeak~\cite{vero_tableak_2023}, as it is the first work on tabular data leakage. The \acrshort{RA} of a reconstructed dataset $\hat{D}$ is computed by averaging the reconstruction accuracy scores of all its samples. This score $ra_{\hat{x}}$ for a single sample $\hat{x} \in \hat{D}$ is defined as the proportion of features for which the inferred value (or range) overlaps, within a specified tolerance, with the corresponding ground-truth value in a paired sample $x \in D$, where $D$ is the original dataset. Formally, $ra_{\hat{x}}$ is computed as follows:
\begin{equation}
  \scriptsize
  ra_{\hat{x}}(x, \hat{x}) = \frac{1}{K+L} \left( \sum_{i=1}^{K} \mathds{1}\bigl(x_i^{(k)} = \hat{x}_i^{(k)}\bigr) + \sum_{i=1}^{L} \mathds{1}\Bigl(\hat{x}_i^{(l)} \in \bigl[x_i^{(l)} - \epsilon_i, x_i^{(l)} + \epsilon_i\bigr]\Bigr) \right)
\end{equation}
where $K$ is the number of categorical features, $L$ is the number of continuous features, and $\epsilon_i$ is the tolerance for the $i$-th continuous feature. Each $\epsilon_i$ is computed by taking the standard deviation of the $i$-th continuous feature in the training dataset and multiplying it by a constant. In our experiments, we follow the setup of TabLeak~\cite{vero_tableak_2023}, and set $\epsilon_i = 0.319 \cdot \sigma_i^{(l)}$, obtaining the following error distribution:
\begin{equation}
    \small
    P\Bigl(\mu - 0.319\cdot\sigma \le x \le \mu + 0.319\cdot\sigma\Bigr) = 2\Phi(0.319) - 1 \approx 25\%,
\end{equation}
where $\Phi$ denotes the cumulative distribution function of the standard normal distribution.
The reconstruction assessment requires each reconstructed sample to be compared with its corresponding ground truth, giving equal weight to all features. To establish a one-to-one correspondence between the reconstructed and original datasets, we employ a bipartite matching algorithm—specifically, the Hungarian method~\cite{kuhn_hungarian_1955}—as done in TabLeak~\cite{vero_tableak_2023}. In \Cref{fig:reconstruction_example} we show an example of reconstruction assessment.  Finally, to visualize the practical impact of high \acrshort{RA}s, Table~\ref{tab:tolerance} presents the tolerance required for a feature in the reconstructed sample to be considered a match with its corresponding original sample.

\mypar{Binary Classification task performance} 
To observe the relationship between overfitting/underfitting and \acrshort{RA}, we show in \Cref{tab:model_performance} how the models perform for each considered dataset, federated gradient boosting implementation, and tree depth. The results indicate that, for the Stroke dataset, the model achieves its best F1-score and \acrshort{AUC} with lower tree depth. This suggests that increasing the tree depth leads to overfitting. In contrast, this trend is not observed for the Pima dataset, which is significantly smaller.

\mypar{Defenses} We also evaluate \name in a scenario where clients use \acrfull{DP} to protect their updates from potential privacy leakage. As discussed in \Cref{sec:motivation}, to the best of our knowledge, this is the only existing method that theoretically mitigates our attack under the given threat model. The authors of FedTree~\cite{li_fedtree_2023} introduce three layers of privacy protection (see \Cref{sub:fltrees_background}), which represent the current \acrshort{SOTA} in federated tree-based systems. However, since our attack reconstructs a dataset from the client side, neither \textit{secure aggregation} nor \acrshort{HE} (protection level $L_1$) mitigates \name, as they only protect clients' updates from server-side privacy threats. Therefore, we evaluate their $L_2$ protection, which incorporates \acrshort{DP}, by varying different \textit{privacy budget values} ($\epsilon$) (more details on $\epsilon$-\acrshort{DP} and its implementation in FedTree are provided in \Cref{appendix:background}). The experiments with \acrshort{DP} pertain only to FedTree, as it is the only framework that implements it for tree-based models across the considered frameworks, to the best of our knowledge. However, since their current implementation is complete only in the vertical \acrshort{FL} scenario, we extend it to the horizontal \acrshort{FL} setting to the best of our ability.

\begin{figure*}[t]
\centering
\subfloat[Stroke - Overall Reconstruction Performance]{\includegraphics[width=0.36\textwidth]{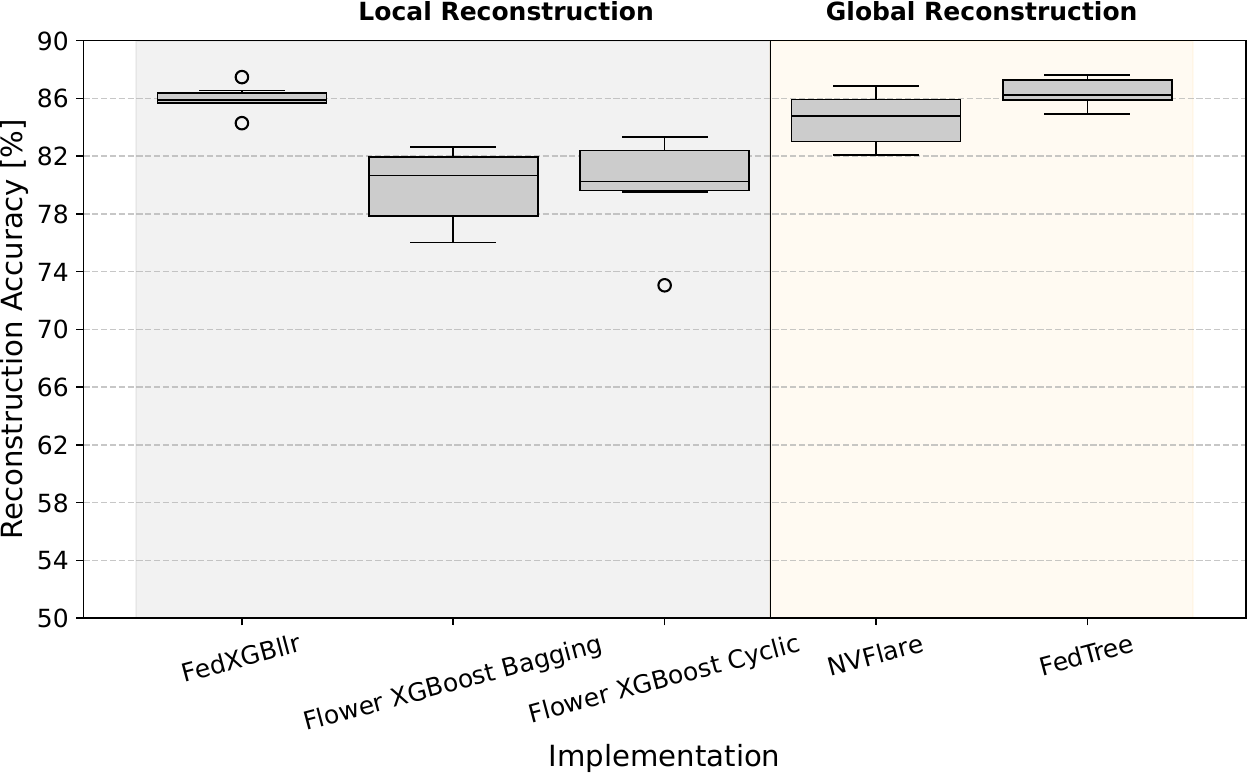}
\label{subfig:stroke_general_perf}}
\hfil
\subfloat[Pima: Overall Reconstruction Performance]{\includegraphics[width=0.36\textwidth]{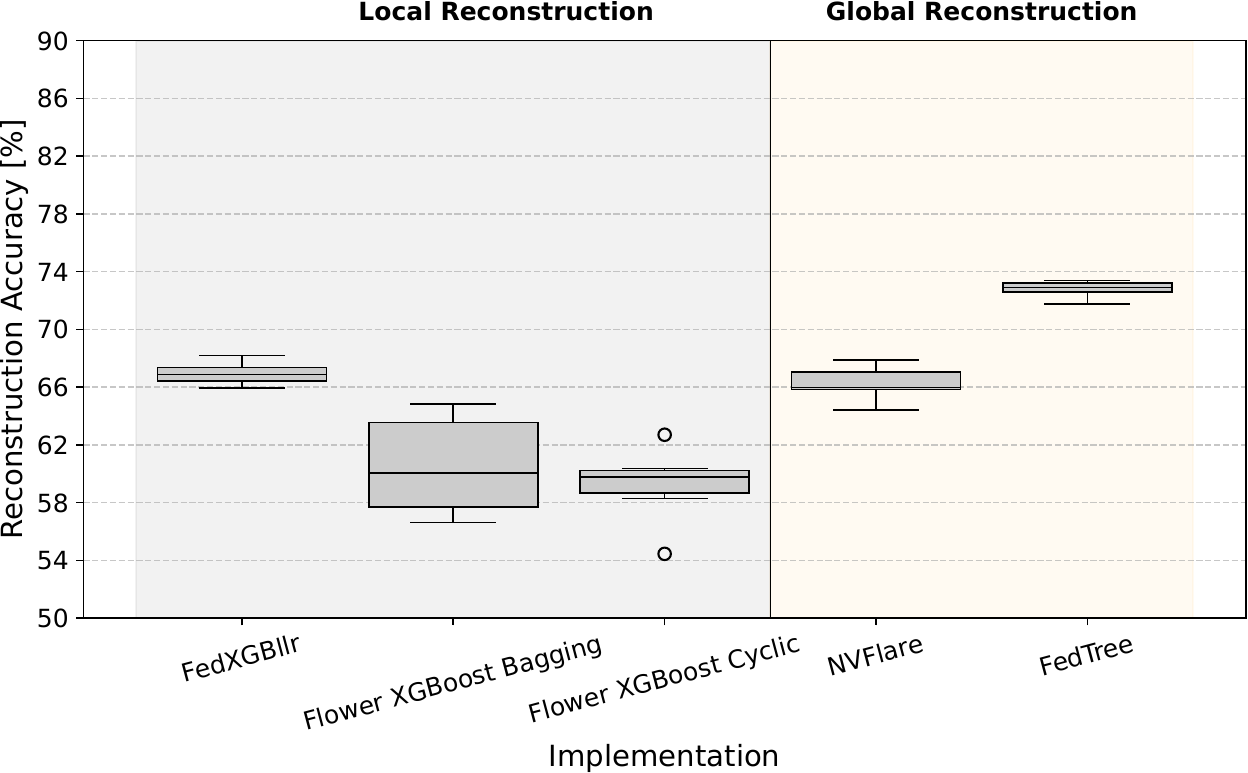}
\label{subfig:pima_general_perf}}
\\
\vspace{-5px}
\subfloat[Stroke: Most Relevant Features Performance]{\includegraphics[width=0.36\textwidth]{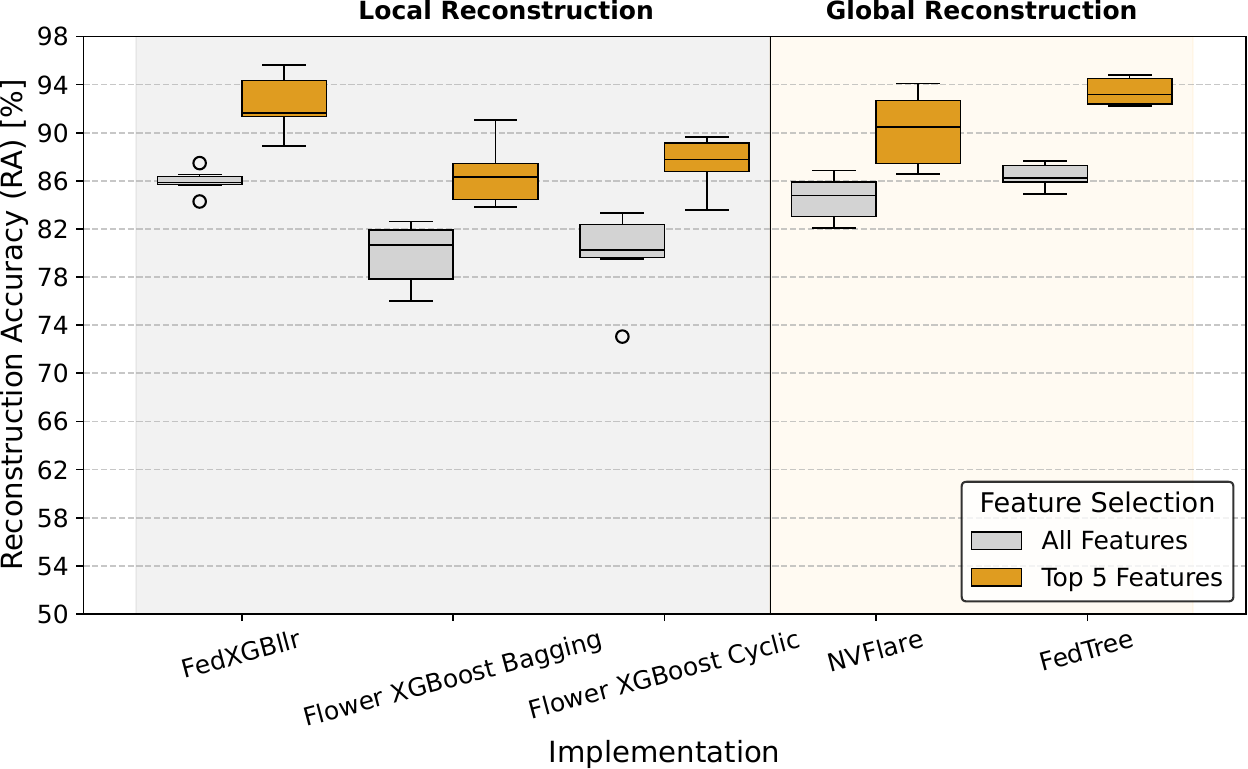}
\label{subfig:stroke_top_perf}}
\hfil
\subfloat[Pima: Most Relevant Features Performance]{\includegraphics[width=0.36\textwidth]{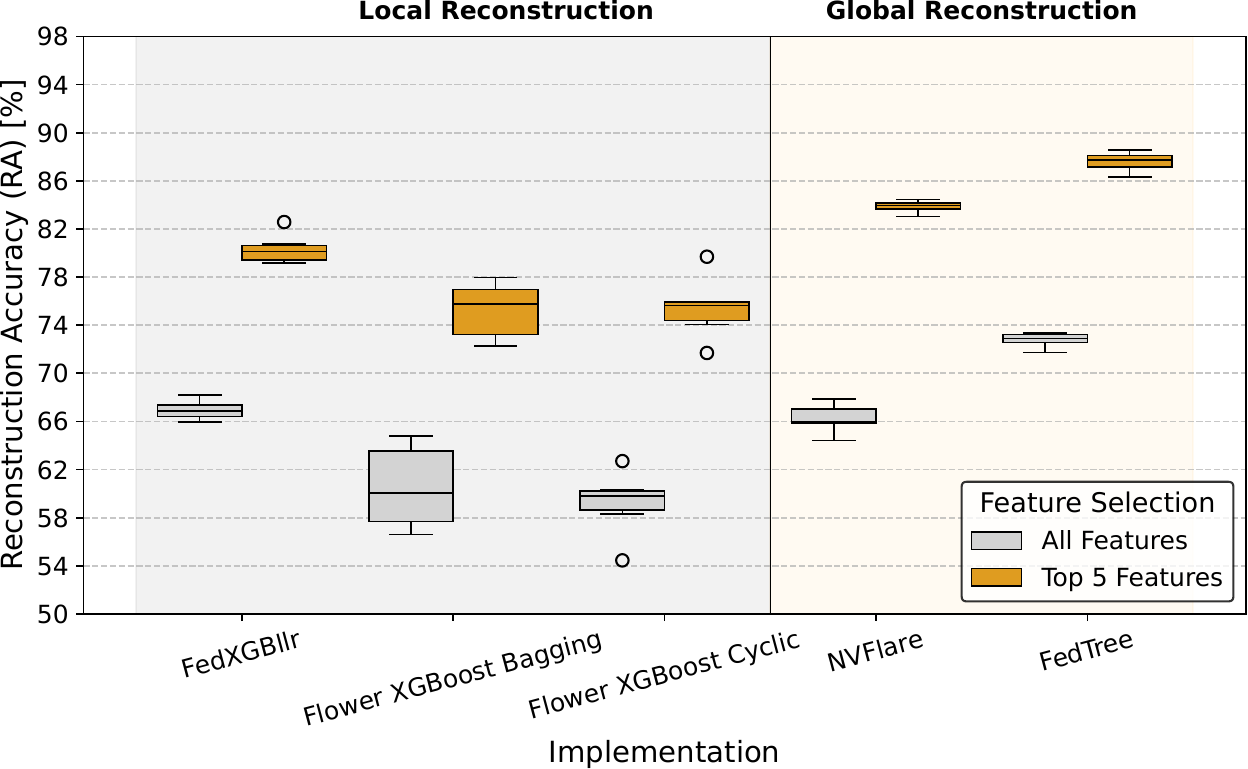}
\label{subfig:pima_top_perf}}
\\
\vspace{-5px}
\subfloat[Stroke: First-Tree Probing (F-TP) \acrshort{RA} vs Final \acrshort{RA}]{\includegraphics[width=0.36\textwidth]{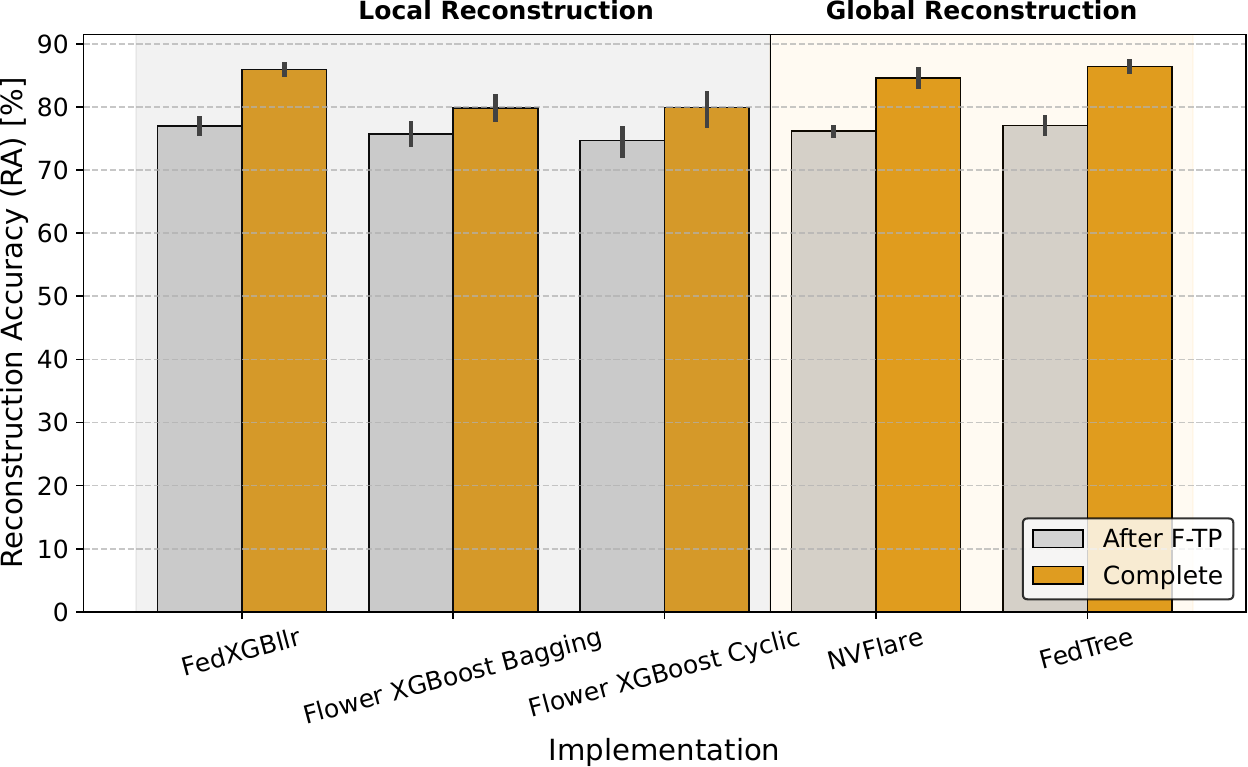}
\label{subfig:stroke_phases_perf}}
\hfil
\subfloat[Pima: First-Tree Probing (F-TP) \acrshort{RA} vs Final \acrshort{RA}]{\includegraphics[width=0.36\textwidth]{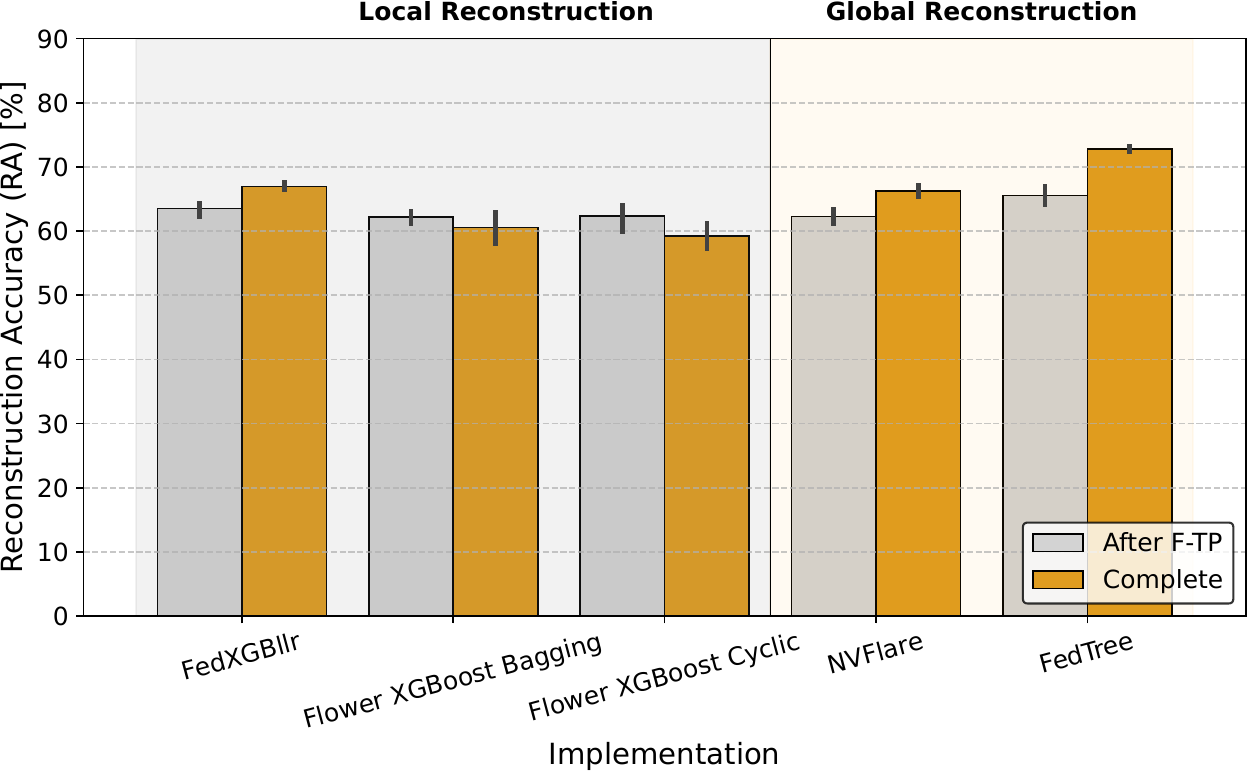}
\label{subfig:pima_phases_perf}}
\\
\vspace{-5px}
\subfloat[Stroke: \acrshort{RA} with Various Tree Depths]{\includegraphics[width=0.36\textwidth]{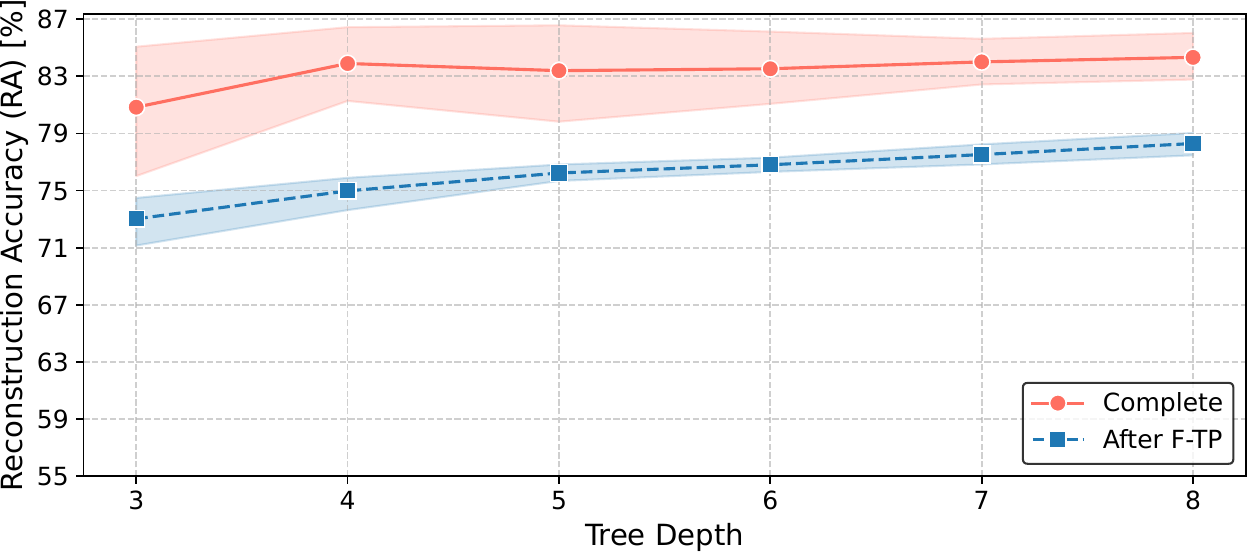}
\label{subfig:stroke_depth_perf}}
\hfil
\subfloat[Pima: \acrshort{RA} with Various Tree Depths]{\includegraphics[width=0.36\textwidth]{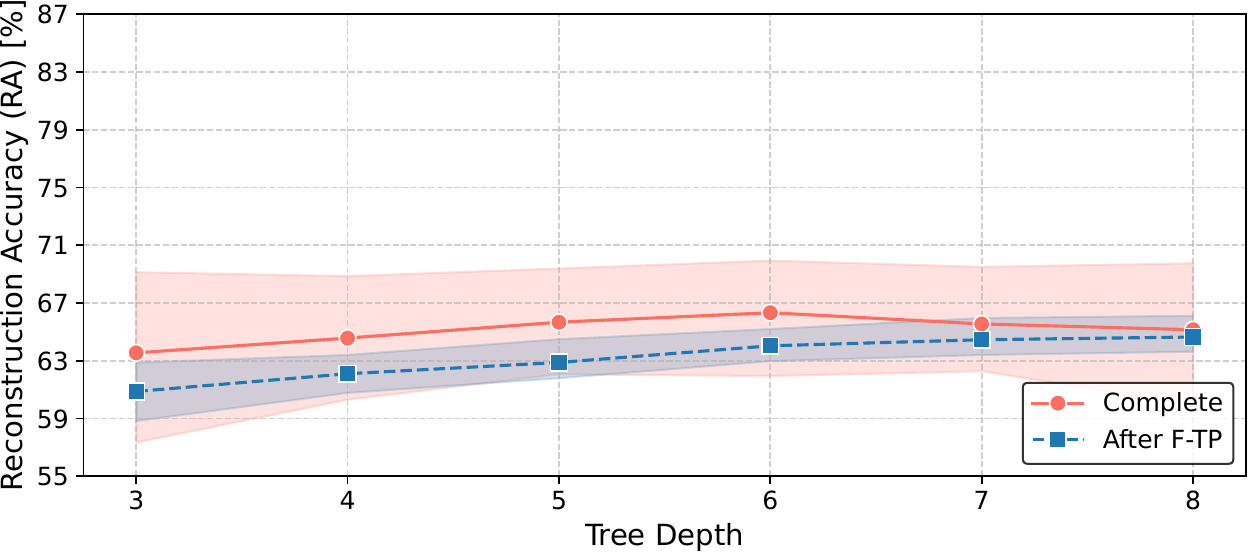}
\label{subfig:pima_depth_perf}}
\vspace{-5px}
\caption{Experimental results on the Stroke Prediction Dataset (left) and the Pima Indians Diabetes Dataset (right).}
\label{fig:combined_perf}
\vspace{-5px}
\end{figure*}

\subsection{\name Effectiveness (RQ1 \& RQ2)} \label{sub:exp_rq1rq2}

We aim to investigate the overall performance of \name, the contribution of its two phases, and how hyperparameters such as tree depth influence the reconstruction performed by our attack.
Following, we present the results using aggregated plots, while more detailed tables with raw results are provided in~\Cref{appendix:detailed_results}. Note that, except for the plots used to show the impact of the client size on \name and the impact of tree depth on \name, all other plots comparing the different implementations in terms of \acrlong{RA} distinguish between \textit{local} and \textit{global} reconstruction, as discussed in \Cref{sub:exp_setup}. In particular, local reconstruction refers to implementations that enable targeting a specific client, while global reconstruction refers to those that allow only the reconstruction of the global dataset.

\subsubsection{Attack performance on different implementations} We present the boxplots of the \acrfull{RA} (in percentage) on both the Stroke (Figure~\ref{subfig:stroke_general_perf}) and Pima (Figure~\ref{subfig:pima_general_perf}) datasets, corresponding to the six runs performed on each system while varying the depth of the trees. Note that the results account for all features in the dataset—11 for the Stroke dataset and 9 for the Pima dataset.
The general performance of \name shows a minimum of 73.05\% \acrshort{RA} on Stroke and 54.45\% \acrshort{RA} on Pima, both registered on Flower XGBoost Cyclic, while achieving a maximum of 86.86\% \acrshort{RA} on Stroke (NVFlare) and 73.37\% \acrshort{RA} on Pima (FedTree), demonstrating that a significant amount of the original dataset can be reconstructed within a tolerance in both datasets. These results are in line with the  \acrshort{SOTA} of \acrshort{RA} on \acrshortpl{ANN} for tabular datasets~\cite{vero_tableak_2023}. However, the \acrshort{RA} on the Pima dataset is worse than what \name obtains on the Stroke one. This discrepancy is due to differences in dataset sizes and the composition of the dataset, which contains only numerical features. This result indicates that reconstructing categorical features is easier than reconstructing numerical ones, a finding also observed by the TabLeak authors.
The two datasets exhibit similar behavior when comparing different systems. In both datasets, the implementations where \name achieves the highest and least skewed \acrshort{RA} are the two histogram-based systems (NVFlare and FedTree) and FedXGBllr. The results obtained with the histogram-based implementations demonstrate that, even if we globally reconstruct the dataset, \name can still achieve high performance. Additionally, FedXGBllr confirms what its implementation suggests: the plain exposure of trees significantly increases the risk of privacy leakage. In contrast, both the Cyclic and Bagging implementations in Flower prove to be more resistant to attacks. This can be attributed to their interleaved nature, where each client continuously trains a new tree based on the trees received from other clients.

\subsubsection{Impact of the feature importance on the \acrshort{RA}} The results already presented demonstrate that \name can effectively reconstruct clients' data with a high accuracy. However, we argue that the importance of features in predictions significantly impacts the \acrshort{RA}. If this holds, it implies that an adversary may achieve even better reconstruction performances on the most relevant features, and may also identify which features they can ``trust'' more in the reconstructed dataset. To validate this intuition, we evaluate the  \acrlong{RA} on the top features (i.e., the most important ones) and compare it to the \acrshort{RA} achieved on the full dataset. Specifically, we train a centralized \acrshort{XGBoost} model, extract the feature importance, and select the top five features (including the label) for each dataset, visible in Table~\ref{tab:tolerance}. In Figures~\ref{subfig:stroke_top_perf} and~\ref{subfig:pima_top_perf}, we compare the \acrshort{RA} evaluated on all features with the one computed using only the top five features. As the figure illustrates, feature importance significantly impacts \name. Indeed, in the best-case scenario, \name achieves an \acrshort{RA} of over 95.63\% for FedXGBllr on the Stroke dataset. Furthermore, \name demonstrates that even with models trained on small and numerical datasets like Pima, on the most relevant (and therefore most impactful on the trees) features, it can still reconstruct well the training data, as shown by the substantial increase in \acrshort{RA}.

\subsubsection{Contribution of the Feature Range Inference phase} In Figures~\ref{subfig:stroke_phases_perf} and~\ref{subfig:pima_phases_perf}, we show the \acrshort{RA} achieved by \name after the first phase (First-Tree Probing) and after the complete attack. As the figure illustrates, the second phase (Feature Range Inference) effectively improves the \acrshort{RA} compared to the first phase alone. In particular, the Feature Range Inference phase increases the average
RA across all systems by 7.19\% on the Stroke dataset and by 1.97\% on the Pima dataset. However, there are two exceptions (Flower Bagging and Cyclic on Pima), where the second phase decreases the average results. Such an exception occurs because the interleaved nature of these implementations, combined with the small dataset size of the Pima dataset, prevents the correct identification of the victim client's tree chain (see \Cref{sub:details_approach}). Interestingly, the \acrshort{RA} achieved using only the first phase demonstrates that the first tree alone is sufficient to cause significant privacy leakage.

\subsubsection{Impact of Client Size on Global Reconstruction}\label{subsub:scalability}
\begin{figure}
    \centering
    \includegraphics[width=0.86\linewidth]{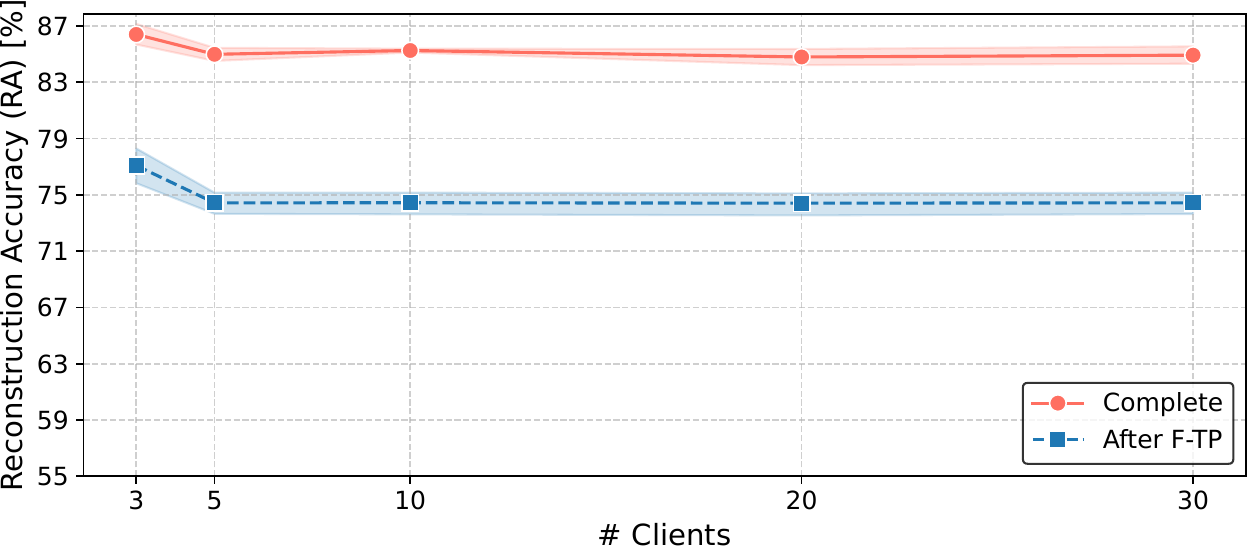}
    \caption{\acrshort{RA} on the Stroke dataset using FedTree, with varying numbers of clients involved in the training.}
    \label{fig:scalability_exp}
\end{figure}

We previously discussed the difference between global and local reconstruction. In the case of the two histogram-based methods, the reconstruction targets the global dataset rather than any single client. Here, we investigate whether increasing the number of participating clients leads to a loss of detail in the aggregated information. To show the impact of client scaling on global reconstruction, we present in \Cref{fig:scalability_exp} the results of an experiment where we evaluate \name against FedTree on the Stroke Dataset in five different scenarios (3, 5, 10, 20, and 30 clients). As the plot depicts, except for a small drop from the scenario with 3 clients and the scenario with 5 clients, the effect of client set size is negligible for the rest of the scenarios, as no significant drop in \acrshort{RA} is registered. Therefore, we can state that the \acrshort{RA} is poorly affected by the number of clients when performing a global reconstruction on a histogram-based system. The theoretical basis for this result derives from the histogram aggregation mechanism. In horizontal \acrshort{FL}, client-side histograms are aggregated by summation. This preserves the characteristics of the individual local histograms, making the resulting global histograms equivalent to the statistics obtained by training a centralized model on the entire dataset. However, as the authors of FedTree~\cite{li_fedtree_2023} point out, when the splits proposed by different clients diverge, they merge the local histograms by approximate summation. Considering that increasing the number of clients leads to a finer partitioning of the dataset, and consequently fewer samples per client, this approximation becomes the source of performance degradation for \name. However, such degradation is negligible, as shown in \Cref{fig:scalability_exp}.

\begin{answer}
\textbf{Answer to RQ1.} \name achieves a \acrlong{RA} comparable to the \acrshort{SOTA} of \acrshort{RA} on \acrshortpl{ANN} for tabular datasets. Additionally, its \acrshort{RA} significantly improves, reaching a maximum of 95.63\%, on the most important features in the dataset. The Feature Range Inference phase increases the average \acrshort{RA} across all systems by 7.19\% compared to First-Tree Probing on the Stroke dataset and by 1.97\% on the Pima dataset. Nonetheless, results demonstrate that First-Tree Probing can achieve good performance even on its own.
\end{answer}

\subsubsection{Impact of the tree depth} When training a gradient boosting model, it is possible to set the tree depth. This hyperparameter is crucial for training performance, and we expect that it also impacts the \acrshort{RA} of our attack. 
In Figures~\ref{subfig:stroke_depth_perf} and~\ref{subfig:pima_depth_perf}, we show the trend of \acrlong{RA}, averaged from the five different implementations, as the tree depth used for training in the ensemble varies. As the plots illustrate, the first phase is significantly impacted by the tree depth. This result is expected: deeper trees contain more splits, allowing us to extract more information from the first tree. Furthermore, this result is in line with the \acrshort{SOTA} of privacy attacks that demonstrate overfitting (demonstrated in~\Cref{tab:model_performance} for Stroke) helps privacy leakage~\cite{securitynet}. In contrast, the overall attack results do not exhibit a consistently increasing trend. This can be attributed to the time constraint imposed on solving the \acrshort{MILP} problem in the Feature Range Inference phase. Higher tree depth increases complexity and, eventually, performance saturates.

\begin{answer}
\textbf{Answer to RQ2.} Increasing tree depth enhances the attack’s effectiveness in the initial phase by providing more splits, which allows for greater information extraction. This supports the intuition that overfitting increases privacy leakage risks. However, as depth increases further, the overall attack does not show a consistently improving trend. This is due to the higher computational complexity in the Feature Range Inference phase and the time constraint, which leads to performance saturation.
\end{answer}

\begin{table}[t]
  \centering
  \caption{Model Utility (F1-score and \acrshort{AUC}), \acrshort{RA} after First-Tree Probing (F-TP) and final \acrshort{RA} under \acrshort{DP}. Both histogram-level and total $\epsilon$ are reported.}
  \label{tab:dp_aggregated_table}
  \resizebox{\linewidth}{!}{%
  \begin{tabular}{ccccccc}
      \toprule
      \textbf{Features} & \textbf{$\epsilon_{\text{histogram}}$} & \textbf{$\epsilon_{\text{total}}$} & \textbf{F1} & \textbf{AUC} & \textbf{RA (F-TP)} & \textbf{RA} \\
      \midrule
      \multirow{4}{*}{All Features}
      & No Defense & --  & 0.601 ± 0.009 & 0.810 ± 0.010 & 65.54 ± 1.95 & 72.78 ± 0.60 \\
      & 1 & 200 & 0.536 ± 0.036 & 0.781 ± 0.033 & 60.65 ± 3.96 & 70.52 ± 1.77 \\
      & 0.25 & 50  & 0.481 ± 0.043 & 0.650 ± 0.037 & 59.97 ± 4.10 & 61.59 ± 3.22 \\
      & 0.125 & 25 & 0.448 ± 0.042 & 0.605 ± 0.055 & 60.26 ± 1.91 & 54.43 ± 4.63 \\
      \midrule
      \multirow{4}{*}{Top 5 Features} 
      & No Defense & --  & 0.601 ± 0.009 & 0.810 ± 0.010 & 82.65 ± 2.03 & 87.59 ± 0.81 \\
      & 1 & 200 & 0.536 ± 0.036 & 0.781 ± 0.033 & 75.95 ± 3.88 & 85.84 ± 1.17 \\
      & 0.25 & 50 & 0.481 ± 0.043 & 0.650 ± 0.037 & 73.26 ± 7.73 & 75.70 ± 4.17 \\
      & 0.125 & 25 & 0.448 ± 0.042 & 0.605 ± 0.055 & 74.18 ± 3.32 & 67.65 ± 5.54 \\
      \bottomrule
  \end{tabular}
  }
\end{table}

\subsection{DP Effectiveness against \name (RQ3)} \label{sub:exp_rq3} We now want to assess if \acrshort{SOTA} defenses mitigate \name. To do so, we test the use of \acrfull{DP} as a defense against \name by using the FedTree implementation.

In this experiment, we use only the Pima dataset, as the Stroke dataset already exhibits poor performance, making it difficult to observe significant changes in model performance. We analyze \name's behavior by varying a fundamental parameter for \acrshort{DP}, i.e., the privacy budget $\epsilon$. In $\epsilon$-\acrshort{DP} (more details in \Cref{appendix:background}), this parameter measures the ``privacy loss,'' meaning that the higher the $\epsilon$, the less privacy the update preserves.

\Cref{tab:dp_aggregated_table} depicts the results obtained by evaluating \name both without defense and with varying values of $\epsilon$, selected according to \acrshort{SOTA} settings. We report both the \emph{histogram-level} $\epsilon$ ($\epsilon_{\text{histogram}}$) and the corresponding \emph{total} $\epsilon$ ($\epsilon_{\text{total}}$). Note that in FedTree, $\epsilon$-\acrshort{DP} is satisfied at the histogram level (details in \Cref{appendix:background}). Unlike \acrshort{ANN}-based federated protocols, here the updates that a client aims to protect are the histograms containing the proposed splits rather than the entire model. Thus, $\epsilon_{\text{histogram}}$ is the effective privacy parameter governing each shared update. $\epsilon_{\text{total}}$ depends on the number of trees and is provided for completeness, but it does not represent the privacy guarantee for any individual communication round.

Analyzing the model's performance, we observe a significant drop even under the least restrictive $\epsilon$. This decline is particularly evident in the F1-score and persists at lower values of $\epsilon$. Overall, compared to the ``no defense'' configuration, we observe a maximum average decrease of 0.153 in F1-score and 0.205 in \acrshort{AUC}, highlighting the impact of \acrshort{DP} on model usability. On the other hand, when examining \acrshort{RA}, we find that the reduction in performance between the ``no defense'' configuration and the least restrictive privacy setting is less pronounced than the utility drop, with an average decrease of 2.26\% when considering all features and 1.75\% when considering only the top five. Additionally, the First-Tree Probing phase of our attack demonstrates to be less affected by \acrshort{DP} than the overall attack, even showing a higher \acrshort{RA} when $\epsilon_{\text{histogram}} = 0.125$.

Overall, while \acrshort{DP} reduces the effectiveness of our attack by sacrificing model utility, it does not fully mitigate \name. Indeed, even under the most aggressive privacy setting and considering the entire feature space, \name can still achieve an \acrshort{RA} $>$ 50\%.

\begin{answer}
\textbf{Answer to RQ3.} \acrlong{DP} reduces the attack’s effectiveness but does not fully mitigate it. Indeed, while \acrshort{DP} lowers the \acrlong{RA}, it also significantly degrades model utility, with average drops of 0.153 in F1-score and 0.205 in AUC in the most aggressive scenario considered. Moreover, even under the strictest privacy setting, the attack still achieves an \acrshort{RA} > 50\%, which means \acrshort{DP} is insufficient as a standalone defense.
\end{answer}
\section{\name Mitigation Guidelines}\label{sec:mitigation}

Our evaluation shows that while \acrshort{DP} reduces the effectiveness of \name, it does not fully mitigate the attack and significantly degrades model utility. Additionally, our \acrshort{SOTA} analysis suggests that other existing defenses fail to address the threat model assumed for \name. Among the implementations considered, histogram-based methods appear to be the most promising direction for limiting information accessible to clients, as they only allow the reconstruction of global rather than local datasets. 

Motivated by our findings, we argue that a tree-based \acrshort{FL} protocol must be designed with a clear understanding of the unique attack surfaces introduced by the federated setting and the model itself. In light of this, we define guidelines for the future design of privacy-preserving federated tree-based systems. Specifically, we describe how histogram-based frameworks like FedTree can be modified to limit client-side visibility of globally aggregated statistics while theoretically preserving the functionality.

\name exploits FedTree's sharing of first-order and second-order gradients ($G$ and $H$). While server-side threats can already be mitigated by existing defenses such as \acrshort{HE} and secure aggregation, mitigating this vulnerability on the client side requires a different approach. To this end, we propose a theoretical restriction on the information broadcast by the server. In particular, the protocol should ensure that the server transmits only the final split decision (or leaf value) to the clients, while keeping the underlying \textit{global} gradient ($G$) and Hessian ($H$) statistics confidential. The training of a node can be refined as follows.

\mypar{Client side} For each feature $a$, each client $i$ computes and sends to the server a local histogram $H^{(i)}_a$, which contains the gradients and Hessian values of the split points proposed by the client $i$.

\mypar{Server side} Upon receiving $H^{(i)}_a$ from each client $i$, the server performs the following steps:
\begin{enumerate}
    \item Aggregates the statistics for each proposed split point by summing the histograms (as proposed in FedTree~\cite{li_fedtree_2023}):
    \[H_a = \sum_{i\in [C]}H^{(i)}_a.\]
    \item Computes the gain \( \mathcal{G}(s) \) for each candidate split using the formula in~\Cref{eq:xgboost_split_gain} and selects the optimal split:
    \[
    s^* = \arg\max_s \mathcal{G}(s),
    \]
    or, if it is a leaf node, compute the \textit{leaf value}.
    \item Broadcasts only the selected split \( s^* \) or \textit{leaf value}.
\end{enumerate}

While we believe that this approach effectively eliminates the information leakage vector exploited by \name—specifically, access to global gradients and Hessians—we leave a detailed empirical evaluation of its impact on federated learning performance to future work.
Moreover, this design may reduce transparency and explainability, as clients no longer have visibility into the rationale behind the split decisions. Investigating this trade-off between privacy and interpretability is also left as future work.

\section{Limitations and Future Work} \label{sec:lim_futurework}

The computational complexity of our attack, particularly during the \emph{Feature Range Inference} phase, is theoretically exponential in the worst-case scenario (see~\Cref{appendix:complexity} for further details). Although in our experiments this complexity is limited by the imposed time constraint on the optimizer, it still represents a limitation.

Furthermore, as previously discussed, histogram-based mechanisms allow each attacker controlling a client to see only aggregated statistics, thereby limiting \name. Indeed, with this specific mechanism, \name can only reconstruct the global training dataset (i.e., the union of all clients' datasets) rather than enabling user-level reconstruction. Additionally, the interleaved nature of the Bagging and Cyclic implementations affects the precision of our reconstruction, as demonstrated by the results.

Our current work represents a first attempt to demonstrate privacy leakage risks in tree-based horizontal \acrshort{FL} systems. Looking ahead, important research directions emerge from both our guidelines on a mitigation strategy and the current limitations of our approach. Therefore, we plan to implement a novel tree-based \acrshort{FL} system, for which we gave an insight in~\Cref{sec:mitigation}.

In addition, future research will explore heuristics to reduce the exponential worst-case complexity and empirically evaluate our attack on tasks beyond binary classification (see~\Cref{appendix:multiclass} for a formalization of the multiclass classification task). Finally, future work could explore how incorporating prior model knowledge affects the success and robustness of adversarial strategies.
\section{Conclusion}
This work demonstrated that tree-based horizontal \acrshort{FL} systems are vulnerable to privacy leakage attacks by introducing \name, a dataset reconstruction attack. Our attack exploits aggregated statistics and tree splits to recover sensitive data from other clients' training sets. Our evaluation across five different \acrshort{SOTA} deployed approaches and two tabular datasets from the healthcare domain showed that \name achieves a \acrfull{RA} comparable to \acrshort{SOTA} dataset reconstruction methods for \glspl{ANN} on tabular datasets. Moreover, while applying \acrlong{DP} during training reduced the \acrshort{RA}, it failed to fully neutralize our attack while significantly compromising model utility. The results underscore that federating standard \acrshort{XGBoost} implementations inherently expose privacy vulnerabilities. We suggest that future protocols should retain the benefits of histogram aggregation while avoiding the transmission of aggregated statistics to clients. Although our approach provided critical insights, its limitations in computational complexity and reconstruction granularity in certain implementations also highlighted opportunities for future research.

\begin{acks}
This project has been partially funded by the Horizon EU project TRUSTroke in the call HORIZON-HLTH-2022-STAYHLTH-01-two-stage under GA No. 101080564. and by the Italian Ministry of University and Research (MUR) under the PRIN 2022 PNRR framework (EU Contribution – Next Generation EU – M. 4,C. 2, I. 1.1), SHIELDED project, ID P2022ZWS82, CUP D53D23016240001.

Generative AI tools such as Grammarly and ChatGPT (GPT-4o) were utilized solely for proofreading and grammar refinement in the preparation of this manuscript. The authors retain full responsibility for the content presented in the final version.
\end{acks}

\bibliographystyle{ACM-Reference-Format}
\bibliography{main}


\appendix

\begin{table*}[t]
  \centering
  \caption{\acrfull{RA} after the First-Tree Probing phase (F-TP) [\%], \acrshort{RA} after the complete attack[\%], F1-score, and \acrfull{AUC} on the test set for the binary classification task on the Stroke Prediction Dataset. Each result corresponds to the tree depth $d$ used during training.}
  \label{tab:no_defense_results_stroke}
  \renewcommand{\arraystretch}{1.2}
  \resizebox{\textwidth}{!}{%
  \begin{tabular}{c c c c c c c c c c c c c c c c c c c c c }
      \toprule
      & \multicolumn{12}{c}{\textbf{Local Reconstruction}} & \multicolumn{8}{c}{\textbf{Global Reconstruction}} \\
      \midrule
      \textbf{Depth} 
      & \multicolumn{4}{c}{\textbf{FedXGBllr}}
      & \multicolumn{4}{c}{\textbf{Flower XGBoost Bagging}}
      & \multicolumn{4}{c}{\textbf{Flower XGBoost Cyclic}}
      & \multicolumn{4}{c}{\textbf{NVFlare}}
      & \multicolumn{4}{c}{\textbf{FedTree}}\\
       & RA (F-TP) & RA & F1 & AUC 
       & RA (F-TP) & RA & F1 & AUC 
       & RA (F-TP) & RA & F1 & AUC 
       & RA (F-TP) & RA & F1 & AUC 
       & RA (F-TP) & RA & F1 & AUC \\
      \midrule
      3 & 74.54 & 84.28 & 0.215 & 0.635 & 72.01 & 76.00 & 0.213 & 0.755 & 69.74 & 73.05 & 0.237 & 0.748 & 74.28 & 85.93 & 0.165 & 0.767 & 74.53 & 84.91 & 0.207 & 0.766 \\
      4 & 75.93 & 85.90 & 0.218 & 0.628 & 74.78 & 80.33 & 0.215 & 0.750 & 72.48 & 79.90 & 0.157 & 0.738 & 75.90 & 86.86 & 0.203 & 0.774 & 75.79 & 86.47 & 0.200 & 0.764 \\
      5 & 76.41 & 87.47 & 0.183 & 0.620 & 75.89 & 77.02 & 0.185 & 0.750 & 75.22 & 80.57 & 0.158 & 0.731 & 76.35 & 85.90 & 0.203 & 0.767 & 77.27 & 85.99 & 0.209 & 0.766 \\
      6 & 77.54 & 85.85 & 0.157 & 0.617 & 76.13 & 80.97 & 0.176 & 0.746 & 76.21 & 79.51 & 0.162 & 0.760 & 76.85 & 83.66 & 0.164 & 0.775 & 77.29 & 87.63 & 0.110 & 0.767 \\
      7 & 78.40 & 86.53 & 0.105 & 0.608 & 76.99 & 82.24 & 0.138 & 0.731 & 76.55 & 83.33 & 0.176 & 0.734 & 77.06 & 82.07 & 0.103 & 0.744 & 78.59 & 85.85 & 0.130 & 0.768 \\
      8 & 79.19 & 85.64 & 0.097 & 0.614 & 78.70 & 82.63 & 0.160 & 0.751 & 77.74 & 82.98 & 0.158 & 0.709 & 76.80 & 82.80 & 0.128 & 0.761 & 79.03 & 87.56 & 0.108 & 0.769 \\
      \bottomrule
  \end{tabular}
  }
\end{table*}

\begin{table*}[t]
  \centering
  \caption{\acrfull{RA} (top 50\% important columns) after the First-Tree Probing phase (F-TP) [\%], \acrshort{RA} (top 50\% important columns) after the complete attack [\%] for the binary classification task on the Stroke Prediction Dataset. Each result corresponds to the tree depth $d$ used during training.}
  \label{tab:no_defense_results_stroke_top_columns}
  \renewcommand{\arraystretch}{1.2}
  \resizebox{0.67\textwidth}{!}{%
  \begin{tabular}{c c c c c c c c c c c }
      \toprule
      & \multicolumn{6}{c}{\textbf{Local Reconstruction}} & \multicolumn{4}{c}{\textbf{Global Reconstruction}} \\
      \midrule
      \textbf{Depth} & \multicolumn{2}{c}{\textbf{FedXGBllr}} & \multicolumn{2}{c}{\textbf{Flower XGBoost Bagging}} & \multicolumn{2}{c}{\textbf{Flower XGBoost Cyclic}} & \multicolumn{2}{c}{\textbf{NVFlare}} & \multicolumn{2}{c}{\textbf{FedTree}}\\
       & RA (F-TP) & RA & RA (F-TP) & RA & RA (F-TP) & RA & RA (F-TP) & RA & RA (F-TP) & RA \\
      \midrule
      3 & 76.50 & 88.89 & 75.95 & 85.74 & 76.02 & 83.55 & 75.90 & 94.10 & 77.15 & 92.23 \\
      4 & 79.10 & 91.34 & 78.65 & 83.83 & 76.63 & 87.13 & 78.29 & 91.55 & 78.71 & 92.33 \\
      5 & 78.52 & 95.63 & 80.42 & 84.00 & 77.79 & 88.40 & 79.11 & 93.05 & 80.42 & 93.84 \\
      6 & 79.59 & 91.95 & 78.82 & 86.92 & 77.46 & 86.66 & 80.20 & 89.42 & 81.99 & 94.71 \\
      7 & 82.64 & 95.15 & 81.25 & 87.59 & 78.98 & 89.63 & 78.80 & 86.79 & 83.38 & 92.49 \\
      8 & 82.72 & 91.34 & 82.95 & 91.05 & 79.45 & 89.38 & 75.70 & 86.57 & 83.74 & 94.80 \\
      \bottomrule
  \end{tabular}
  }
\end{table*}

\begin{table*}[t]
  \centering
  \caption{\acrfull{RA} after the First-Tree Probing phase (F-TP) [\%], \acrshort{RA} after the complete attack [\%], F1-score, and \acrfull{AUC} on the test set for the binary classification task on the Pima Indians Diabetes Dataset. Each result corresponds to the tree depth $d$ used during training.}
  \label{tab:no_defense_results_diabetes}
  \renewcommand{\arraystretch}{1.2}
  \resizebox{\textwidth}{!}{%
  \begin{tabular}{c c c c c c c c c c c c c c c c c c c c c }
      \toprule
      & \multicolumn{12}{c}{\textbf{Local Reconstruction}} & \multicolumn{8}{c}{\textbf{Global Reconstruction}} \\
      \midrule
      \textbf{Depth} & \multicolumn{4}{c}{\textbf{FedXGBllr}} & \multicolumn{4}{c}{\textbf{Flower XGBoost Bagging}} & \multicolumn{4}{c}{\textbf{Flower XGBoost Cyclic}} & \multicolumn{4}{c}{\textbf{NVFlare}} & \multicolumn{4}{c}{\textbf{FedTree}}\\
       & RA (F-TP) & RA & F1 & AUC & RA (F-TP) & RA & F1 & AUC & RA (F-TP) & RA & F1 & AUC & RA (F-TP) & RA & F1 & AUC & RA (F-TP) & RA & F1 & AUC \\
      \midrule
      3 & 63.82 & 68.20 & 0.587 & 0.816 & 59.85 & 57.56 & 0.638 & 0.769 & 57.07 & 54.45 & 0.573 & 0.753 & 60.57 & 65.81 & 0.643 & 0.817 & 63.08 & 71.74 & 0.586 & 0.810 \\
      4 & 60.57 & 66.36 & 0.603 & 0.806 & 62.34 & 58.02 & 0.587 & 0.743 & 64.20 & 59.74 & 0.506 & 0.669 & 60.20 & 66.01 & 0.676 & 0.834 & 63.21 & 72.74 & 0.596 & 0.826 \\
      5 & 63.00 & 66.61 & 0.572 & 0.810 & 61.55 & 62.05 & 0.577 & 0.716 & 61.93 & 59.82 & 0.490 & 0.669 & 61.97 & 67.38 & 0.639 & 0.821 & 65.98 & 72.51 & 0.606 & 0.798 \\
      6 & 64.70 & 67.42 & 0.531 & 0.806 & 62.71 & 64.81 & 0.506 & 0.689 & 62.83 & 58.29 & 0.591 & 0.755 & 63.79 & 67.87 & 0.636 & 0.831 & 66.16 & 73.27 & 0.604 & 0.805 \\
      7 & 64.52 & 65.94 & 0.607 & 0.809 & 63.32 & 64.03 & 0.537 & 0.703 & 63.85 & 60.35 & 0.549 & 0.719 & 63.26 & 64.41 & 0.627 & 0.832 & 67.35 & 73.03 & 0.604 & 0.806 \\
      8 & 64.24 & 67.14 & 0.582 & 0.814 & 63.32 & 56.61 & 0.585 & 0.718 & 64.38 & 62.70 & 0.529 & 0.679 & 63.84 & 65.94 & 0.660 & 0.825 & 67.46 & 73.37 & 0.610 & 0.815 \\
      \bottomrule
  \end{tabular}
  }
\end{table*}

\begin{table*}[t]
  \centering
  \caption{\acrfull{RA} (top 50\% important columns) after the First-Tree Probing phase (F-TP) [\%], \acrshort{RA} (top 50\% important columns) after the complete attack [\%] for the binary classification task on the Pima Indians Diabetes Dataset. Each result corresponds to the tree depth $d$ used during training.}
  \label{tab:no_defense_results_diabetes_top_columns}
  \renewcommand{\arraystretch}{1.2}
  \resizebox{0.67\textwidth}{!}{%
  \begin{tabular}{c c c c c c c c c c c }
      \toprule
      & \multicolumn{6}{c}{\textbf{Local Reconstruction}} & \multicolumn{4}{c}{\textbf{Global Reconstruction}} \\
      \midrule
      \textbf{Depth} & \multicolumn{2}{c}{\textbf{FedXGBllr}} & \multicolumn{2}{c}{\textbf{Flower XGBoost Bagging}} & \multicolumn{2}{c}{\textbf{Flower XGBoost Cyclic}} & \multicolumn{2}{c}{\textbf{NVFlare}} & \multicolumn{2}{c}{\textbf{FedTree}} \\
       & RA (F-TP) & RA & RA (F-TP) & RA & RA (F-TP) & RA & RA (F-TP) & RA & RA (F-TP) & RA \\
      \midrule
      3 & 78.39 & 82.57 & 78.26 & 72.67 & 71.55 & 71.69 & 75.62 & 84.20 & 79.67 & 87.87 \\
      4 & 77.44 & 79.21 & 79.37 & 74.91 & 78.56 & 75.86 & 78.04 & 83.82 & 80.67 & 86.32 \\
      5 & 80.39 & 80.74 & 79.68 & 77.97 & 80.36 & 75.44 & 79.10 & 83.03 & 83.71 & 87.57 \\
      6 & 79.68 & 79.98 & 79.80 & 76.62 & 79.09 & 74.04 & 80.62 & 84.07 & 82.95 & 88.17 \\
      7 & 79.74 & 79.15 & 80.33 & 77.09 & 80.33 & 75.97 & 81.13 & 83.58 & 84.07 & 87.02 \\
      8 & 79.27 & 80.27 & 81.18 & 72.26 & 80.39 & 79.69 & 83.14 & 84.45 & 84.80 & 88.57 \\
      \bottomrule
  \end{tabular}
  }
\end{table*}

\begin{table*}[t]
  \centering
  \scriptsize
  \setlength{\tabcolsep}{4pt}
  \renewcommand{\arraystretch}{1.2}
  \caption{FedTree with \acrfull{DP} on different values of privacy budget $\epsilon$ (Both
histogram-level and total $\epsilon$ are reported). \acrfull{RA} after the First-Tree Probing phase (F-TP) [\%], \acrshort{RA} after the complete attack [\%], F1-score, and \acrfull{AUC} on the test set for the binary classification task on the Pima Indians Diabetes Dataset. Each result corresponds to the tree depth $d$ used during training.}
  \label{tab:defense_results_diabetes}
  \begin{tabular}{c cccc cccc cccc cccc}
      \toprule
      \textbf{Depth} & \multicolumn{4}{c}{\textbf{No Defense}} & \multicolumn{4}{c}{$\epsilon_{\text{histogram}} = 1$ | $\epsilon_{\text{total}} = 200$} & \multicolumn{4}{c}{$\epsilon_{\text{histogram}} = 0.25$ | $\epsilon_{\text{total}} = 50$} & \multicolumn{4}{c}{$\epsilon_{\text{histogram}} = 0.125$ | $\epsilon_{\text{total}} = 25$} \\
      \cmidrule(lr){2-5}
      \cmidrule(lr){6-9}
      \cmidrule(lr){10-13}
      \cmidrule(lr){14-17}
        & RA (F-TP) & RA & F1 & AUC & RA (F-TP) & RA & F1 & AUC & RA (F-TP) & RA & F1 & AUC & RA (F-TP) & RA & F1 & AUC \\
      \midrule
      3 & 63.08 & 71.74 & 0.586 & 0.810 & 54.32 & 67.48 & 0.500 & 0.811 & 53.88 & 57.43 & 0.471 & 0.645 & 56.47 & 48.31 & 0.495 & 0.694 \\
      4 & 63.21 & 72.74 & 0.596 & 0.826 & 58.84 & 69.41 & 0.589 & 0.819 & 57.02 & 61.89 & 0.415 & 0.614 & 60.75 & 50.75 & 0.481 & 0.641 \\
      5 & 65.98 & 72.51 & 0.606 & 0.798 & 59.92 & 71.14 & 0.568 & 0.798 & 60.37 & 58.53 & 0.547 & 0.708 & 60.54 & 55.94 & 0.472 & 0.599 \\
      6 & 66.16 & 73.27 & 0.604 & 0.805 & 62.39 & 71.38 & 0.531 & 0.739 & 59.90 & 61.64 & 0.468 & 0.681 & 61.56 & 52.82 & 0.386 & 0.589 \\
      7 & 67.35 & 73.03 & 0.604 & 0.806 & 62.43 & 71.29 & 0.529 & 0.769 & 63.94 & 64.12 & 0.496 & 0.626 & 61.55 & 58.18 & 0.417 & 0.555 \\
      8 & 67.46 & 73.37 & 0.610 & 0.815 & 65.99 & 72.39 & 0.500 & 0.751 & 64.69 & 65.93 & 0.487 & 0.624 & 60.70 & 60.55 & 0.435 & 0.550 \\
      \bottomrule
  \end{tabular}
\end{table*}

\begin{table*}[t]
  \centering
  \scriptsize
  \setlength{\tabcolsep}{4pt}
  \renewcommand{\arraystretch}{1.2}
  \caption{FedTree with \acrfull{DP} on different values of privacy budget $\epsilon$ (Both
histogram-level and total $\epsilon$ are reported). \acrfull{RA} (top 50\% important columns) after the First-Tree Probing phase (F-TP) [\%], \acrshort{RA} (top 50\% important columns) after the complete attack [\%] for the binary classification task on the Pima Indians Diabetes Dataset. Each result corresponds to the tree depth $d$ used during training.}
  \label{tab:defense_results_diabetes_top_columns}
  \resizebox{0.7\textwidth}{!}{%
  \begin{tabular}{c cccc cccc cccc cccc}
      \toprule
      \textbf{Depth} & \multicolumn{2}{c}{\textbf{No Defense}} & \multicolumn{2}{c}{$\epsilon_{\text{histogram}} = 1$ | $\epsilon_{\text{total}} = 200$} & \multicolumn{2}{c}{$\epsilon_{\text{histogram}} = 0.25$ | $\epsilon_{\text{total}} = 50$} & \multicolumn{2}{c}{$\epsilon_{\text{histogram}} = 0.125$ | $\epsilon_{\text{total}} = 25$} \\
      \cmidrule(lr){2-5}
      \cmidrule(lr){6-9}
      \cmidrule(lr){10-13}
      \cmidrule(lr){14-17}
        & RA (F-TP) & RA & RA (F-TP) & RA & RA (F-TP) & RA & RA (F-TP) & RA\\
      \midrule
      3 & 79.67 & 87.87 & 72.01 & 84.77 & 59.69 & 72.37 & 68.05 & 60.21\\
      4 & 80.67 & 86.32 & 71.55 & 84.12 & 69.73 & 75.95 & 74.00 & 65.93\\
      5 & 83.71 & 87.57 & 74.62 & 86.24 & 74.59 & 70.30 & 74.02 & 70.30\\
      6 & 82.95 & 88.17 & 78.20 & 87.02 & 75.43 & 75.05 & 77.74 & 63.22\\
      7 & 84.07 & 87.02 & 77.93 & 85.97 & 78.77 & 78.75 & 76.17 & 70.98\\
      8 & 84.80 & 88.57 & 81.38 & 86.92 & 81.35 & 81.76 & 75.08 & 75.24\\
      \bottomrule
  \end{tabular}
  }
\end{table*}

\section{Ethical Considerations}\label{appendix:ethical}

This research was conducted with a clear commitment to ethical responsibility, carefully considering the impact on all relevant stakeholders. The decisions made throughout the study balanced transparency, security, and long-term benefits. 
We are aware that the publication of attacks may raise ethical concerns, as it exposes techniques that could be exploited by malicious actors. However, failing to address these risks would leave users unaware of real-world threats. By presenting our research, we highlight the privacy risks of tree-based \acrlong{FL} while also providing mitigation guidelines, including insights into the limitations of differential privacy and other defenses, which may not be as effective as users assume. This approach ensures that users can take informed security measures. Responsible disclosure of threats, paired with the provision of practical countermeasures, helps prevent adversaries from gaining an asymmetric advantage. Our work follows this principle by equipping the community with both an understanding of potential risks and the means to mitigate them.

\section{Background}\label{appendix:background}

In this section, we complete the background by providing details about \acrfull{DP}.

\subsection{Differential Privacy}
\acrfull{DP}~\cite{dwork_differential_2006,ji_differential_2014} is a privacy definition that ensures that the output of a computation does not reveal too much information about any individual in the dataset.
The idea is to add noise to the output of the computation in such a way that the privacy of the individuals is preserved.
Formally, a randomized algorithm $\mathcal{M}$ satisfies $\epsilon$-differential privacy if for all datasets $D$ and $D'$ that differ in one element, and for all subsets of the output space $S$:
\begin{equation}
    \label{eq:dp_definition}
    \Pr[\mathcal{M}(D) \in S] \leq e^{\epsilon} \Pr[\mathcal{M}(D') \in S].
\end{equation}
The parameter $\epsilon$, also called \textit{privacy budget}, is a measure of the privacy loss; the smaller the value of $\epsilon$, the more privacy is preserved.
The Laplace mechanism is a simple way to achieve \acrshort{DP} by adding Laplace noise to the output of the computation.
The Laplace mechanism is defined as:
\begin{equation}
    \label{eq:laplace_mechanism}
    \mathcal{M}(D) = f(D) + \text{Lap}(\frac{\Delta f}{\epsilon}),
\end{equation}
where $f(D)$ is the output of the computation on the dataset $D$, $\Delta f$ is the sensitivity of the function $f$, and $\text{Lap}(\lambda)$ is the Laplace distribution with scale $\lambda$.
\subsubsection{FedTree $\epsilon$-DP implementation}

The FedTree~\cite{li_fedtree_2023} protocol implies the sharing of histograms at each round. Therefore, it implements $\epsilon$-\acrshort{DP} at the level of histogram sharing. Unlike \acrshort{FL} protocols based on \acrlongpl{ANN}, where the gradients or model weights are the objects of protection, in FedTree, the sensitive information resides in the \emph{histograms}, which contain the aggregated gradient and Hessian statistics used to construct tree nodes.

To satisfy $\epsilon$-\acrshort{DP} for each shared histogram, FedTree follows the procedure we explain below.

\mypar{Gradient Clipping} First-order gradients and the second-order ones are clipped with a threshold of  $R$ and $2R$ respectively, to ensure bounded sensitivity.

\mypar{Laplace Noise Addition} Laplace noise is added to each element of the histogram. Specifically, the noise is drawn from the Laplace distribution with mean $0$ and scale $\frac{2R}{\epsilon}$, i.e., $\text{Lap}(0, \frac{2R}{\epsilon})$. This guarantees that each shared histogram satisfies $\epsilon$-\acrshort{DP}.

\mypar{Histogram-Level Privacy} \acrshort{DP} is applied \textit{per histogram}, meaning that the privacy budget $\epsilon_{\text{histogram}}$ governs the noise added to each update (i.e., histograms), rather than to the entire model.

\mypar{Total Privacy Budget} In their implementation\footnote{\url{https://github.com/Xtra-Computing/FedTree}}, FedTree allows setting an aggregated privacy budget $\epsilon_{\text{total}}$, which represents the upper bound on the cumulative privacy loss across all trees. As a result, this budget depends on the number of trees. Specifically, given a value for $\epsilon_{\text{total}}$, the privacy budget allocated to each leaf node histogram, $\epsilon_{\text{histogram}}$, is computed using the following formula:
\begin{equation}
\epsilon_{\text{histogram}} = \frac{\epsilon_{\text{total}}}{2T}
\end{equation}
where $T$ denotes the number of trees in the model.

\section{Detailed Results}\label{appendix:detailed_results}

In this section, we provide the raw results of experiments in Sections~\ref{sub:exp_rq1rq2} and~\ref{sub:exp_rq3}. Specifically, the results regarding~\Cref{sub:exp_rq1rq2} for the Stroke Dataset are presented in Tables \ref{tab:no_defense_results_stroke} and~\ref{tab:no_defense_results_stroke_top_columns}, while the results for the Diabetes Dataset are shown in Tables~\ref{tab:no_defense_results_diabetes} and~\ref{tab:no_defense_results_diabetes_top_columns}. Finally, the results regarding~\Cref{sub:exp_rq3} are presented in Tables~\ref{tab:defense_results_diabetes} and~\ref{tab:defense_results_diabetes_top_columns}.

\section{Attack Computational Complexity}\label{appendix:complexity}
In this section, we analyze the computational complexity of our proposed reconstruction attack. As shown in \Cref{sec:approach}, our attack consists of two phases: \emph{First-Tree Probing} and \emph{Feature Range Inference}. In the following, we theoretically analyze the complexity of each phase.

\mypar{First-Tree Probing} In this phase, intermediate values are computed for each leaf, requiring \(\mathcal{O}(L)\) operations, where \(L\) denotes the number of leaves in the \acrshort{DT} and \(d\) its depth. Additionally, for each leaf, the \acrshort{DT} is traversed to aggregate the corresponding feature range information. Since each traversal requires \(\mathcal{O}(\log d)\) time, processing all \(L\) leaves results in an additional cost of \(\mathcal{O}(L \cdot \log d)\). Therefore, the overall computational complexity of the \emph{First-Tree Probing} phase is \(\mathcal{O}(L \cdot \log d)\).

\mypar{Feature Range Inference} This phase involves solving a series of \acrfull{MILP} problems for binary decision-making. In general, \acrshort{MILP} is NP-hard and has a worst-case computational complexity that is exponential in the number of binary decision variables. In our formulation, each variable \(x_{ij}\) indicates whether sample \(i\) is assigned to leaf \(j\). If every sample between the $n$ samples in the dataset could be assigned to any of the \(m\) leaves, the search space would contain \(\mathcal{O}(n \cdot m)\) binary variables, leading to exponential complexity in the worst case.  However, our model introduces constraints of the form \(x_{ij} = 0\) for all \(j \notin L_i\), where \(L_i \subseteq J\) denotes the subset of leaves that sample \(i\) can reach, based on inferred feature ranges. As the attack progresses through successive trees, these feature ranges become increasingly constrained, reducing the size of \(L_i\) for each sample. This prunes the effective search space, significantly mitigating the theoretical worst-case complexity in practical scenarios. To further control optimization time, during our experimental evaluation, we impose a $10$-minute time constraint on the \acrshort{MILP} solver for each tree. This ensures that the optimization phase remains computationally feasible, even when the worst-case complexity is prohibitive, while having a minimal impact on the overall reconstruction quality.

\section{Extension to Multiclass Classification}\label{appendix:multiclass}
In \Cref{sec:approach}, we provided a formalization of our attack against federated gradient boosting binary classifiers. In this section, we extend the theoretical formalization to multiclass classification by highlighting the differences with the binary case.

\acrfull{GBDT}~\cite{friedman_greedy_2001} and \acrshort{XGBoost}~\cite{chen_xgboost_2016} approach the multiclass classification task using a \acrfull{OvR} strategy~\cite{ovr_paper}. In this approach, each boosting iteration trains $K$ trees—one for each of the $K$ classes—where each tree is trained to distinguish one class from all the others. Unlike the binary classification case, which typically uses the log loss function, the multiclass setting employs a softmax loss function. This allows each class-specific tree to be optimized by considering the predictions from all $K$ class-specific trees from the previous boosting iteration.

Regarding federated implementations of multiclass algorithms, their mechanisms are equivalent to those used for binary classification. Therefore, the information available on the client side remains the same as discussed in the binary case.

We now describe the two phases of \name (\textit{First-Tree Probing} and \textit{Feature Range Inference}) adapted to the multiclass setting.

\subsection{First-Tree Probing}
As in the binary classification case, the controlled client targets a specific client and analyzes the first trees trained by that client. In the multiclass scenario, this corresponds to analyzing the first $K$ trees, one for each of the $K$ classes. Each of these trees distinguishes between class 1 (the class $c$ for which the tree is trained) and class 0 (all other classes).

The base score in the multiclass case differs from that in the binary case. Indeed, while in binary classification it is directly interpreted as a probability, in the multiclass setting the base score acts as a logit and is the same for all $K$ classes. Therefore, the initial probability for each class is:
\[
p_i^{(c)} = \frac{1}{K}, \quad \forall c \in \{1, \dots, K\}.
\]

As discussed in \Cref{sub:fist-tree-prob}, the \textit{First-Tree Probing} phase consists of four steps, which we now describe in the multiclass context.

\mypar{Inferring the Number of Samples per Leaf} 
After clients receive the aggregated trees from the \acrshort{PS}, the adversary can infer the number of samples assigned to each leaf by analyzing the first $K$ trees trained by the victim client. In the multiclass setting, the loss function is the softmax cross-entropy, and we use its derivatives~\cite{chen15, Goodfellow-et-al-2016} to compute the per-sample gradient and Hessian statistics.

For a sample $i$ and class $c$, the gradient and Hessian value are:
\begin{align}
    g_i^{(c)} = p_i^{(c)} - \mathds{1}[y_i = c],   \quad   h_i^{(c)} = 2 \cdot p_i^{(c)} \cdot (1 - p_i^{(c)}),
\label{eq:hessian_gradient_probability_multiclass}
\end{align}
where $p_i^{(c)}$ is the softmax probability for class $c$.
The total Hessian $H^{(c)}_j$ in a leaf $j$ of the first tree for class $c$ can then be expressed as:
\begin{equation}
    H^{(c)}_j = \sum_{i=1}^{N_j} h_{ij}^{(c)} = N_j \cdot 2 \cdot p_i^{(c)} \cdot (1 - p_i^{(c)}),
\end{equation}
and solving for $N_j$ (the number of samples in the leaf) we get:
\begin{equation}
    N_j = \frac{H_j^{(c)}}{2 \cdot p_i^{(c)} \cdot (1 - p_i^{(c)})}.
\end{equation}

\mypar{Inferring the Label Distribution in the Leaves}
After inferring the number of samples $N_j$ in a leaf $j$, we recover the label distribution using the gradient for each leaf $j$ and class $c$:
\begin{equation}
    G^{(c)}_j = -\frac{leaf\_value_j^{(c)}}{\eta} \cdot (H^{(c)}_j + \lambda),
\end{equation}
where $\eta$ is the learning rate and $\lambda$ is the regularization parameter.

Using the per-class aggregated gradient expression, we compute the number of samples $N_j^{(c)}$ for class $c$ in the leaf $j$ as:
\begin{equation}
    N^{(c)}_j = N_j \cdot p_i^{(c)} - G^{(c)}_j.
\end{equation}
Computing this for all $K$ classes (i.e., $K$ trees), we obtain the label distribution in each leaf, where 1 represents the class $c$ for which the tree has been trained, while 0 represents each other class.

\mypar{Dataset Initialization}
At this point, the adversary knows the number of samples per leaf and their class distribution. As in the binary case, they can initialize a dataset by placing the appropriate number of samples with the inferred class labels into each leaf. Tree paths are again used to constrain feature ranges, and samples are initialized uniformly within those ranges. 

However, in the multiclass setting, the ``path traversal'' used to initialize feature ranges is not performed for every sample in the first \( K \) trees. Instead, for each client-specific tree corresponding to class \( c \), only the samples with inferred label \( y_i = c \) are initialized and assigned feature constraints.

\mypar{Compute per-sample Statistics}
Finally, for each inferred sample and class, the adversary computes the probability score $p_i^{(c)}$ with the softmax function:
\begin{equation}
    p_i^{(c)} = \text{softmax} \left( base\_score + \sum_{t=1}^{T} leaf\_value_i^{(c, t)} \right),
\end{equation}
where $\text{leaf\_value}_i^{(c, t)}$ is the value of the leaf reached by sample $i$ in the $t$-th tree for class $c$, and $T$ is the number of boosting rounds.

The gradients and Hessian values are then computed using Equations~\eqref{eq:hessian_gradient_probability_multiclass}, and are used in the second phase of the attack.

\subsection{Feature Range Inference}
In the multiclass setting, the second phase of \name proceeds similarly to the binary case. Since each boosting iteration generates \( K \) class-specific trees (one per class), the adversary solves one optimization problem for each of these trees. However, thanks to the \acrshort{OvR} strategy, each tree can be treated as a binary classifier that distinguishes class \( c \) from the rest. Therefore, the same \acrfull{MILP} formalization described for the binary case can be applied directly, provided that the appropriate values of \( p_i^{(c)} \), \( g_i^{(c)} \), and \( h_i^{(c)} \) are used for each client-specific tree corresponding to class \( c \). In addition, since the client-specific trees are built in parallel during training, the per-sample statistics can be updated only after both the optimization problem and the feature range update have been completed for each of the $K$ trees in the same boosting iteration.

As in the binary case, this iterative refinement improves the accuracy of the reconstructed dataset by refining the feature ranges based on the per-tree leaf assignments.

\end{document}